\documentstyle[epsf]{article}

\newcommand{\bee}{\begin{equation}}
\newcommand{\ee}{\end{equation}}
\newcommand{\beea}{\begin{eqnarray}}
\newcommand{\eea}{\end{eqnarray}}
\newcommand{\rme}{{\rm e}}
\newcommand{\ewxy}[2]{\setlength{\epsfxsize}{#2}\epsfbox[10 60 640 570]{#1}}

\begin{document}
\thispagestyle{empty}
\parskip=12pt
\raggedbottom

\def\mytoday#1{{ } \ifcase\month \or
 January\or February\or March\or April\or May\or June\or
 July\or August\or September\or October\or November\or December\fi
 \space \number\year}
\noindent
\hspace*{9cm} COLO-HEP-397\\
\vspace*{1cm}
\begin{center}
{\LARGE Tests of Hypercubic Fermion Actions}

\vspace{0.5cm}

T. DeGrand\\
Physics Department, 
        University of Colorado, \\ 
        Boulder, CO 80309 USA

(the MILC collaboration)

\begin{abstract}
I have performed scaling tests using quenched spectroscopy
 of a family of 
fermion actions which have a hypercubic kinetic term, gauge connections
built of smeared links, and an anomalous magnetic moment term.
These actions show improved rotational invariance compared to the standard
Wilson action and to the  tadpole-improved clover action.
Hyperfine splittings are improved compared to the standard Wilson action
(at the level of a factor of three in the lattice spacing),
and are  about the same as for the tadpole-improved  clover action.
\end{abstract}

\end{center}
\eject

\section{Introduction}
I report on tests of a family of fermion actions for lattice gauge
theory simulations, which are designed to improve scaling of hadron
spectroscopy.  While very expensive to simulate, they appear to more
than repay their computational cost with better scaling behavior.

The actions which I tested were inspired by the fixed point (FP)
action program \cite{FP,FP2}  for 
fermions\cite{Wiese,BW,FPF,MITFP1,MITFP2,LANG,FARCHIONI,PETER97}.
However, they are not FP actions. They
 are  Wilson-fermion like, in the sense that they have four
component spinors on all sites and there
are no manifest symmetries which protect the bare quark mass from 
being additively renormalized.

The new features which I tested include:

1) a hypercubic kinetic energy term. Each fermion in the action communicates
with $3^4-1=80$ nearest neighbors.  This term improves the hadron
dispersion relation compared to that from actions using a
 standard on-axis nearest neighbor coupling.

2) Gauge connections built of very fat gauge links, links built by 
 averaging
the fundamental link variables over a local region
and re-projecting them
 back onto the gauge group.
This construction removes short distance fluctuations from the 
correlators during the simulation process, rather than attempting
to divide them out at the end.  This results in observed very small additive
mass renormalization of bare quantities, and I conjecture that all
perturbative corrections to observables are considerably reduced.
To include very fat links in a simulation with dynamical fermions using
known technology might be quite expensive, but in quenched simulations
the cost is minimal.

I also tested a complicated lattice anomalous magnetic momentum term in which
the quark and antiquark are not fixed to the same site.
Some kind of  term is needed to correct the lattice free 
quark magnetic moment and to improve meson and baryon hyperfine splittings.
However, my tests show that the standard ``clover'' term, suitably
normalized, improves scaling as much  as the complicated Pauli term
needed to satisfy the FP equations.

The scaling tests in this paper are a bit non-standard (when compared to
other studies of spectroscopy) and deserve some explanation.
I am interested in scaling tests which are uncontaminated by
extrapolations in volume or to the chiral limit.  Thus I compare only
simulations in fixed physical volume.  Any volume would do, and so
I choose a small one simply because these actions are expensive to
simulate.  To set the scale I use a gluonic observable because these
are quenched simulations. As a choice of gluonic observables,
one has $T_c$, the critical temperature for deconfinement,  the string
tension $\sigma$ or the Sommer radius $r_0$. Glueball or torelon measurements
are just too costly. Of these observables, the ones associated with the
potential ($\sigma$ and $r_0$) require a fit to a function $V(r)$;
 the choice of the
fitting function can affect the results. At very coarse lattice spacing
this problem becomes more serious.  Thus I use $T_c$ to
set the scale, and do simulations on lattices
of fixed size $L=2/T_c$. With $\sqrt{\sigma} \simeq 440$ MeV 
and $\sqrt{\sigma}/T_c \simeq 1.60$, the scales are $T_c=275$ MeV,
$L=1.45$ fm, and $2\pi/L=860$ MeV.  Lattice spacings $aT_c=1/2$, 1/3, 1/4
correspond to $a=0.36$, 0.24, and 0.18 fm.

All the tests are performed at fixed physical quark mass (defined either
by interpolating lattice data to a fixed value of $m_\pi/m_\rho$
or to a fixed value of $m_\pi/T_c$).  At very coarse
lattice spacing, and with heavy (though still relativistic) quarks,
 scaling violations from conventional actions
 are very large. Thus only modest statistics are required to identify
improvement--or lack thereof--compared to them.

Besides spectroscopy, I
measure the meson or baryon dispersion relation.
In lattices of fixed physical volume set by $T_c$,
 the physical momenta corresponding
to the different allowed lattice modes are multiples of $T_c$,
$a\vec p = 2\pi \vec n/L$ or $\vec p = \pi T_c \vec n$, if $L=2/T_c$,
and one can compare data
with different lattice spacings  at the same physical momentum.
Wilson and clover fermions at $\beta<6.0$ (using the Wilson gauge action)
exhibit bad scaling or rotational invariance violations. 
 The new actions are rotationally
invariant even at $a=0.36$ fm.

The outline of the paper is as follows:
In Section 2 I describe the  new features of these actions.
Section 3 is devoted to scaling tests, and I make some tentative conclusions in
Section 4.
I describe the (new) FP gauge action used in these simulations in the Appendix.

\section{Ingredients of the Actions}

\subsection{Hypercubic kinetic term}
The fermionic free field action has the generic form
\bee
\Delta_0(x) = \lambda(x) + i\sum_\mu \gamma_\mu \rho_\mu(x).
\ee
It is constructed by finding some free FP action \cite{Wiese,BW,FPF,MITFP1}.
We begin with a continuum action for  fermions
which has no doublers and is chirally symmetric.
 We construct an
action on a coarser distance scale $\Delta'$ by iterating the FP
equation
\beea
(\Delta')^{-1}_{n_b,n'_b} =  & {1\over \kappa} \delta_{n_b,n'_b} \nonumber \\
+  & \Omega_{n_b,n} (\Delta)^{-1}_{n,n'}\Omega_{n',n'_b}^T \nonumber \\
\label{RGF}
\eea
where $\Omega$ is a blocking kernel and $\kappa$ is a tunable parameter.
We  select
a blocking kernel, iterate the RGT to find a fixed point action,
and then tune parameters in $\Omega$ to make the action maximally local.
We have used a factor-of-two rescaling
 in which $\Omega$ is restricted to a hypercube:
$\Omega_{ij}$ is nonzero only if
$j=i\pm\mu$, $i\pm\mu\pm\nu, \dots$ $i\pm\mu\pm\nu\pm\lambda\pm\sigma$,
and $\Omega_{ij}=c((i-j)_1,(i-j)_2,(i-j)_3,(i-j)_4)$.
Each site communicates to $3^4-1=80$ neighbors.

 There are many good parameterizations, resulting in
fairly local FP actions.  However,
one ultimately wants to use these actions in simulations, and the
action must be somehow truncated. There are a number of (subjective) criteria
to select a good RGT, based on the properties of the truncated action
(which the RGT does not know about): a good dispersion relation,
$E(p) = |\vec p|$ out to large $|\vec p|$ with no complex roots, good
free-field thermodynamics, $P = 1/3 \sigma T^4$ even at large
discretization, etc.
We tuned the RGT to optimize these criteria, and our choice
is given by $c(1,0,0,0)=0.03$, $c(1,1,0,0)=0.01$,
 $c(1,1,1,0)=0.005$, $c(1,1,1,1)=0.0025$, 
and $\kappa=44.0$.

\begin{figure}[htb]
\begin{center}
\vskip -10mm
\leavevmode
\epsfxsize=60mm
\epsfbox[40 50 530 590]{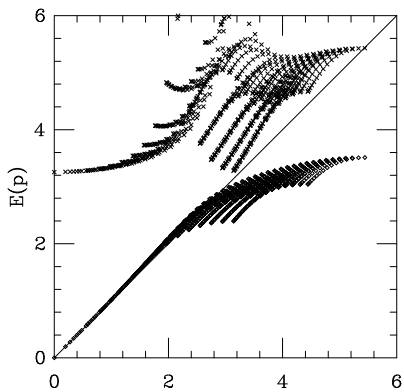}
\vskip -5mm
\end{center}
\caption{Dispersion relation $E(p)$ vs $|p|$ for $m_0=0$ for the hypercubic  action.}
\label{fig:disp1}
\end{figure}

\begin{figure}[htb]
\begin{center}
\vskip -10mm
\leavevmode
\epsfxsize=60mm
\epsfbox[40 50 530 590]{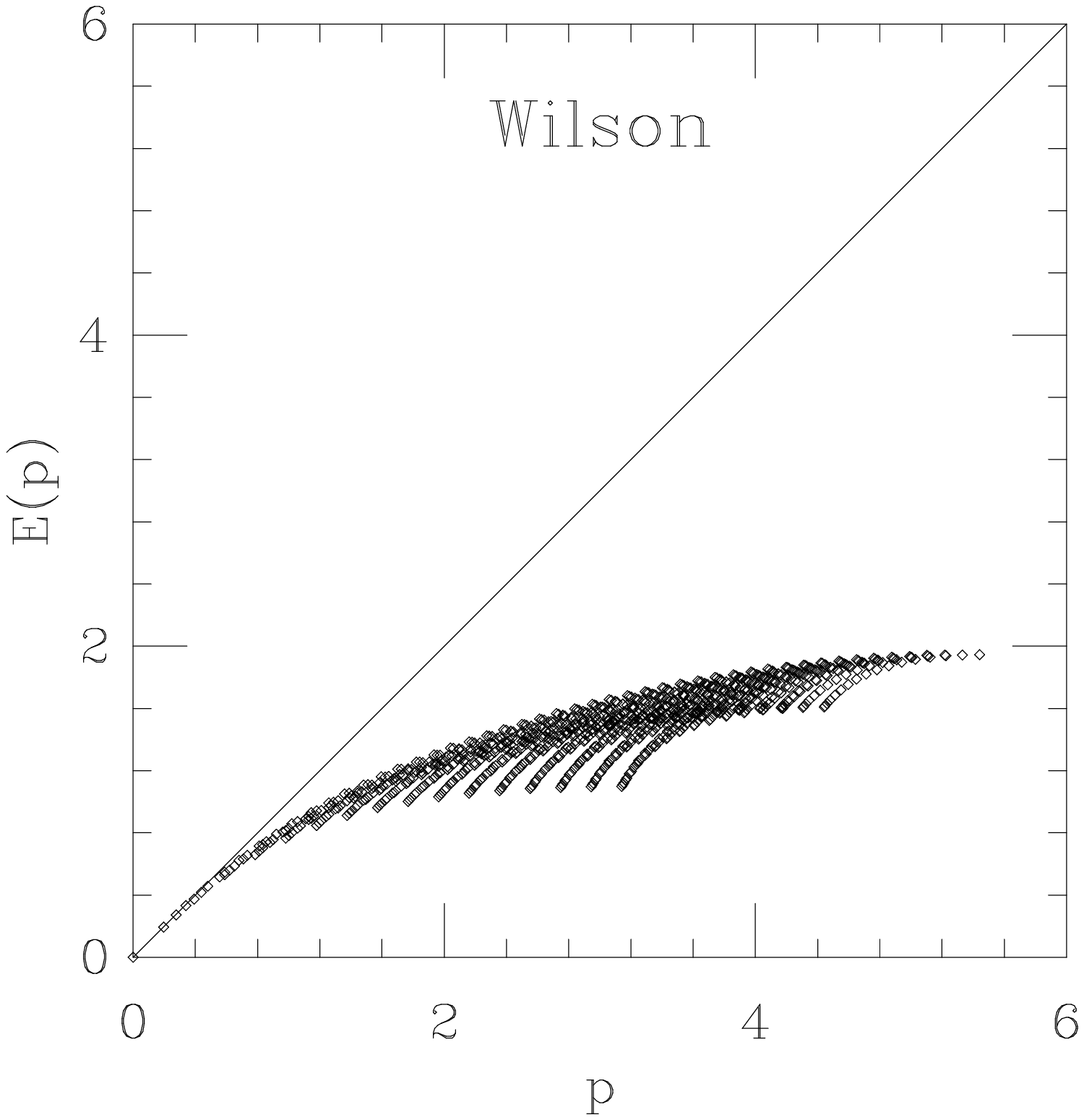}
\vskip -5mm
\end{center}
\caption{Dispersion relation $E(p)$ vs $|p|$ for massless Wilson fermions.}
\label{fig:wilsondr}
\end{figure}

The couplings of the FP action for our  RGT (for massless fermions) fall
off exponentially with $r=\sqrt{\sum_\mu x_\mu^2}$.
The largest entries at distance $r=2$
are at location $x=(\pm 1,\pm 1,\pm 1,\pm 1)$.  The smallest
truncation which 
accurately reproduces the main features of this FP action
 is to an action which
sits on a hypercube--that is, the
free field action is to be parameterized 
with five nonzero $\lambda$'s and four nonzero $\rho$'s, corresponding
to each of the nonzero offsets.
An example of a dispersion relation for this hypercubic
action is compared to the Wilson action in Figs. \ref{fig:disp1} and 
\ref{fig:wilsondr}.
We show both branches of the hypercubic action's dispersion relation;
all roots are real.
The non-truncated FP action has a perfect dispersion relation $E= |\vec p|$
for all $ \vec p$.

For massive fermions, we need an action which is on a renormalized trajectory
(RT)
for some RGT (with  the mass scaling at each step with the
ratio of lattice spacings).
 To reach the RT, one can begin with
 an action which has a very small mass
but is otherwise close to a FP action, perform a series of blockings,
and follow it out.

One complication with this procedure is that an action which is local for
small mass can block into an action for large mass which is not short range.
To avoid this, I take an RG transformation whose parameters are functions
of the mass and tune the
 parameter(s) to insure a local
action at each blocking step. The resulting $\lambda$'s and $\rho$'s
are smooth functions of the  mass.
Again, the dispersion relation for hypercubic approximations to
RT actions are well behaved out to large $|\vec p|$.

Plots of the variation of the
 parameters--or the tables of numbers corresponding to them--are
 by themselves not very useful for calculation. Rather than give them,
I will immediately present a simple parameterization of the couplings.

My choice of parameterization is to regard all the $\rho$'s and $\lambda's$
as linear functions of the bare mass and to determine $\lambda(0,0,0,0)$
for positive bare mass by solving the dispersion relation for it:
defining
\bee
R= 2\rho_0(1,0,0,0)+12\rho_0(1,1,0,0)+24\rho_0(1,1,1,0)+16\rho_0(1,1,1,1)
\ee
\bee
A_s=2\lambda(1,0,0,0)+12\lambda(1,1,0,0)+24\lambda(1,1,1,0)+16\lambda(1,1,1,1)
\ee
\bee
D=8\lambda(1,0,0,0)+24\lambda(1,1,0,0)+32\lambda(1,1,1,0)+16\lambda(1,1,1,1)
\ee
the pole of the propagator ($\lambda^2(p) + \sum_\mu \rho_\mu^2(p)=0$
at $p_\mu=(im_0,0,0,0)$) is at
\bee
\lambda(0,0,0,0) = -A_s\cosh(m_0-1) -R\sinh(m_0) - D.
\label{POLE}
\ee
My linear
parameterization runs over $0<m_0<0.32$, which is the useful range
of mass for light hadron spectroscopy for lattice spacings $aT_c=1/2$ to 1/4.

The big problem 
in parameterizing an approximate action
is that the additive renormalization of the quark mass
for $g^2 \neq 0$ drives the critical quark mass $m^c_q$ negative, outside
the region where the solution of the FP equation is meaningful.
With Wilson fermions, and one free parameter ($\kappa$) this is not a problem
(one just tunes $\kappa$ above $\kappa_c$ without
 otherwise altering the action) but with these complicated actions there is 
not a clear cut way to proceed. In principle, all the parameters in the
action should depend on  properties of the gauge field 
(for example, on the local value of the plaquette).
To circumvent this, I make an arbitrary choice:  I assume that the parameters
continue to vary linearly with the bare mass, and I determine $\lambda(0,0,0,0)$
by continuing the low-mass limit of Eqn. \ref{POLE} to negative mass.
\bee
\lambda(0,0,0,0) = -A_s  - D -R m_0.
\label{NEGPOLE}
\ee
Since  $m^c_q \simeq -0.4$ at $\beta_c(N_t=4)$ and moves toward
zero at bigger $\beta$'s,  $\lambda(0,0,0,0)$ is basically a linear function
of $m_0$.

This parameterization is not appropriate for studying charm with this
action (where $am_0=2$ to 4 depending on the lattice spacing).
The parameters of the action on the RT are smooth functions of $m_0$,
so one could probably construct a more complicated (polynomial
or exponential) fit to them.

Table \ref{tab:hyper} gives the parameters of the linear fit.

\begin{table*}[hbt]
\setlength{\tabcolsep}{1.5pc}
\caption{Linear parameterization of the couplings of the hypercubic action:
$\lambda(x_0,x_1,x_2,x_3)= \lambda_a + m_0 \lambda_b$,
$\rho_0(x_0,x_1,x_2,x_3)= \rho_a + m_0 \rho_b$.}
\label{tab:hyper}
\begin{tabular*}{\textwidth}{@{}l@{\extracolsep{\fill}}lcccccc}
\hline
offset   & $\lambda_a$   & $\lambda_b$ & $\rho_a$ & $\rho_b$  \\
\hline
0 0 0 0 & 2.256756  &-0.9863  &  &  \\
1 0 0 0 &  -0.1122  & 0.0741  & -0.1464  & 0.1300   \\
1 1 0 0 &  -0.0323  & 0.0271  & -0.0329  & 0.0303   \\
1 1 1 0 &  -0.0144  & 0.0141  & -0.0101  & 0.0096   \\
1 1 1 1 &  -0.0072  & 0.0076  & -0.0035  & 0.0033   \\
\hline
\end{tabular*}
\end{table*}

\subsection{Very Fat Links}
Measurements of pure gauge observables (the potential or glueball masses)
suffer from noise arising from the short distance fluctuations of the
gauge fields. A good cure for this problem has been known for many years:
define new link variables which do not couple to the UV sector of the
lattice variables, and which have the same IR properties as the original 
variables.  An  example of such a variable is an APE-blocked
link \cite{APEBlock} 
\begin{eqnarray}
V^{n+1}_\mu(x) = (1-\alpha)V^{n}_\mu(x) & +  & \alpha/6 \sum_{\nu \ne \mu}
(V^{n}_\nu(x)V^{n}_\mu(x+\hat \nu)V^{n}_\mu(x+\hat \nu)^\dagger
\nonumber  \\
& + & V^{n}_\nu(x- \hat \nu)^\dagger
 V^{n}_\mu(x- \hat \nu)V^{n}_\mu(x - \hat \nu +\hat \mu) )
\end{eqnarray}
(with $V^0_\mu(x)=U_\mu(x)$ and $V^{n+1}_\mu(x)$ is projected back onto $SU(3)$).  It is also known that for best results, both $\alpha$
and the maximum number of blocking steps $N$ should increase as the lattice
spacing decreases.

Fermions also suffer from bad UV behavior, and their symptoms include the
breaking of flavor symmetry (for staggered fermions), large additive
renormalization of the bare mass (for Wilson fermions), and large
renormalizations of currents (for any kind of fermion).
The tadpole improvement program \cite{LM}
was originally designed to estimate or compute these
large UV effects and subtract  (or divide)
them out during the conversion from the
lattice calculation to  continuum number. 

However, recent evidence
suggests that it may in some cases be better to remove the UV fluctuations
directly
from the simulations.  This evidence is the partial restoration of
flavor symmetry breaking for staggered fermions
by replacing the link by an $N=1$ APE-blocked
link, as shown by Ref. \cite{MILC} and (with a slightly different averaging)
by Ref. \cite{SINCLAIR}.  These authors restrict themselves to $N=1$,
 presumably because they wish to use their actions for simulations 
with dynamical fermions.  However, if one is interested in quenched simulations,
one can APE-block to any desired level, with tiny overhead, simply
by pre-computing and storing the APE-blocked links.  Then if $N>1$
improves UV behavior, one is free to use it.

It is easy to understand why the fat links suppress UV fluctuations
\cite{LEPAGE}.
Each term in the action in coordinate space
\bee
{\cal L}_I = {1\over{2a}}\sum_{x,y,z} \bar \psi(x) \Gamma
\dots U_\mu(x+y) \dots \psi(x+z)
\ee
can be expanded as a power series in $g$
\beea
{\cal L}_I = & {1\over{2a}}\sum_{x,y,z} \bar \psi(x) \Gamma
\dots 
(1 + iga A_\mu(x+y+a/2\hat \mu)  \nonumber \\
 & -{1\over 2}(ag)^2 A_\mu(x+y+a/2\hat \mu)^2  \dots) \dots \psi(x+z)
\nonumber \\
\eea
which in momentum space  
becomes ${\cal L}_I = {\cal L}_I^1 + {\cal L}_I^2$ with
\bee
{\cal L}^1_I = i{g \over 2} \int_{p,q} \bar \psi(p)\Gamma
\int_k \delta^4(k+q-p) A_\mu(k)e^{i(y+a/2\hat \mu)k} \psi(q)e^{iqz}
\ee
\beea
{\cal L}^2_I  = & {g^2a\over{2}}\int_{p,q} \bar \psi(p)\Gamma
\int_{k_1,k_2} \delta^4(k_1+k_2+q-p) A_\mu(k_1) A_\mu(k_2)
\nonumber \\
 & e^{i(y+a/2\hat \mu)(k_1+k_2)} \psi(q)e^{iqz}.
\nonumber \\
\eea
Smearing the link over a distance $r_0$ makes the replacement
\bee
A_\mu(r+ {a\over 2} \hat \mu) \rightarrow \sum_{\mu,\nu} \sum_w
h_{\mu\nu}(w) A_\nu(r+w+{a\over 2} \hat \nu)
\ee
or
\bee
A_\mu(k) \rightarrow \sum_{\mu,\nu}
H_{\mu\nu}(k) A_\nu(k)e^{i(y+a/2(\hat \nu -\hat \mu)k}
\ee
(and a similar formula for ${\cal L}^2_I$)
where the form factor is
$H_{\mu\nu}(k) = \sum_r h_{\mu\nu}(r) e^{ikr}$. Essentially any
 smearing function
suppresses the vertex at $k > \pi/r_0$.  In the language of Ref. \cite{LM},
tadpoles contribute beyond their naive strength
because the UV  divergence of the gluon loop compensates
for the $a$-dependence of the vertex; smearing suppresses the coupling
of the fermion to high momentum gluons.

\subsection{Nonlocal Pauli Term}
Lattice fermions have a magnetic moment which is anomalously small due
to lattice artifacts.
One can parameterize the
vertex through the interaction of a fermion with an infinitesimal
magnetic field $B$: the pole in the propagator 
will be at $E= m_0 + B/2m_B$, where $m_B$ is the so-called magnetic
mass.
We write the momentum-space interaction term as
\bee
 \bar \psi(p)  i \Delta^1_\mu(p,-p')A_\mu(p-p') \psi(p')
\ee
and expand the vertex in Dirac space as
\beea
\Delta^1_\mu(k,-p)= & f_{\mu,0}(k,-p) + f_{\mu,\nu}(k,-p)\gamma_\nu
 + \sum_{\rho<\nu} f_{\mu,\rho\nu}(k,-p) \gamma_\rho \gamma_\nu
\nonumber \\
& + f_{\mu,5}(k,-p)\gamma_5  + f_{\mu, \nu 5}(k,-p)\gamma_\nu \gamma_5
\nonumber \\
\eea
(with an identical labeling for the decomposition in coordinate space).
A lot of algebra \cite{FPF} gives
\bee
m_B = - {{\lambda(im_0)[ \lambda'(im_0) - \rho_0'(im_0)  ] }
\over {\rho_2'(im_0) f_{1,1}(im_0) - i c^1_{12}(im_0) \lambda(im_0)   }}
\label{MB}
\ee
where
\bee
\lambda(im) = \sum_n \rme ^{mn_0} \lambda(n)
\ee
\bee
\lambda'(im) = \sum_n n_0 \rme ^{mn_0} \lambda(n)
\ee
\bee
\rho_0'(im) = -i \sum_n n_0 \rme ^{mn_0} \rho_0(n)
\ee\bee
\rho_2'(im) = -i \sum_n n_2 \rme ^{mn_0} \rho_2(n)
\ee
\bee
f_{1,1}(im) = \sum_{xy} \rme ^{m x} f_{1,1}(x,y)
\ee
\bee
ic^1_{12}(im) = \sum_{xy}(x-2y)_2 \rme ^{m x} f_{1,12}(x,y)
\ee
are all real.

All approximate FP vertices I have seen have a complicated Pauli term
with sizable contributions when the quark and antiquark do not sit
on the same lattice site. As an example, Table \ref{tab:c113m}
shows the contribution of various fermion offsets to $ic^1_{12}(x,im)$
for a vertex based on the RGT we have been using in this paper.
The normalization appears to be reasonably well saturated by fermion
offsets over a cube. The gauge connections are very complicated.

However, it is not clear how important the FP version of the Pauli 
term will be in spectroscopy.  I therefore studied three possibilities:

1) No Pauli term at all. This turns out to give hyperfine interactions
 which are too small.

2)  Keep only the on-site part of the Pauli term (the standard clover term)
but choose its normalization so that $m_B=m_0$.  This is not a FP action.
The gauge links will be fattened like the rest of the links in the action.

3) Restrict the Pauli term to offsets which span a cube. For each offset,
sum over all the minimum-length paths (with their sign factors)
which contribute to the Pauli term. Choose the relative normalization
of the terms to match the FP vertex and the overall normalization to
fix $m_B=m_0$.  Fatten the links if necessary.  This choice
(hereafter called a ``full Pauli action'') is a much better
approximation to a FP action than the second choice (``clover action''),
but if the clover action performs as well in a test, it is the action of choice.

\begin{table}
\begin{tabular}{|c|l|l|l|l|}
\hline
x &   $m_0=0.08$ & $m_0=0.16$ & $m_0=0.32$&   $m_0=0.64$ \\
\hline
0 0 0 0 & -0.0913  & -0.0840  & -0.0709  & -0.0527 \\
1 0 0 0 & -0.2490  & -0.2269  & -0.1883  & -0.1278     \\
1 1 0 0 & -0.2980  & -0.2710  & -0.2253  & -0.1479     \\
1 1 1 0 & -0.1641  & -0.1493  & -0.1248  & -0.0816 \\
1 1 1 1 & -0.0318  & -0.0291  & -0.0246  & -0.0167 \\
\hline
Sum in hypercube:    & -0.8343  & -0.7604 & -0.6340 & -0.4269  \\
\hline
Total: & -0.8974 & -0.8182 & -0.6837 & -0.4807 \\
\hline
\end{tabular}
\caption{Nonlocality of the  Pauli term $ic^1_{12}(x,im)$
for actions along the RT.}
\label{tab:c113m}
\end{table}

For $m_0>0$ the constraint $m_B=m_0$ fixes the normalization of the Pauli term.
I find that the normalization varies roughly linearly with the
bare quark mass.  I choose (arbitrarily) to keep the same linear
dependence with $m_0$ even for negative bare mass.  This would be equivalent
in the standard clover action, to making the size of the
clover term a function of
the hopping parameter $\kappa$, rather than a function of $\beta$.
From a practical point of view the difference is slight: as one varies
the gauge coupling in a simulation, the value of bare quark corresponding
to a particular physical hadron mass shifts, becoming (typically) more negative
as $\beta$ decreases. The input coefficient of the clover term becomes
larger as $\beta$ decreases, so the net result is that one can as well
say that the clover term tracks the bare quark mass, as to say that the
it tracks the bare coupling.
At zero gauge coupling FP actions do not have a mass-independent
Pauli normalization and so the standard practice of making the
normalization mass independent is unnatural. Presumably one could do
simulations with a standard action, 
such as the Wilson-plus-clover action,
tuning $m_B$ to equal $m_0$, although
it is hard to see the point of doing this as long as the dispersion
relation is imperfect.

The linear parameterization reproduces $m_B=m_0$ to within
five per cent for $m_0<0.4$.  For the clover hypercubic action I
could have simply set $m_B=m_0$ for $m_0>0$ by inverting Eqn. \ref{MB},
although I did not do that.

\section{Scaling Tests}

\subsection{Survey of Actions Tested}

Most of the quenched spectroscopy has been done using a new parameterization
of a FP gauge action for SU(3).
In the Appendix I tabulate the critical temperature, string tension, and
Sommer parameters for the gluonic action, so the reader can convert 
to his favorite scaling variable.
One test has been done using the original parameterization of the
action presented in Ref. \cite{SEPT}, and some tests use
 the  SU(3) gauge action of our
recent work on instantons\cite{INST}.
We have made rough measurements of its $\beta_c(N_t)$ for deconfinement
to set the scale.

All the fermionic actions are made gauge-invariant by replacing the 
offsets by an
average over the shortest distance gauge paths.  For example,
\bee
\bar \psi(x) \psi(x+\hat \mu + \hat \nu) \rightarrow
{1\over 2}\bar\psi(x)[V_\mu(x)V_\nu(x+\hat \nu) +V_\nu(x)V_\mu(x+\hat\mu)]
\psi(x+\hat \mu + \hat \nu)
\ee
where $V_\mu(x)$ is either one of the original links or an APE-blocked link.

The cost of a hypercubic action per iteration step during matrix inversion
is about 20 times as expensive as the usual Wilson action, since there
are more neighbors and the Dirac connections are not projectors.
Actions with the complicated Pauli term are about
 56 times as expensive as the usual Wilson action.
 All the  gauge connections are pre-computed, so there are startup and 
storage costs, as well.  I used the stabilized biconjugate gradient (biCGstab)
algorithm for matrix inversion \cite{BICG}.

The two actions which were tested most extensively
 both have a hypercubic kinetic term and APE-blocked links with
$N=7$ and $\alpha=0.3$.  Action A has a full Pauli term. 
It is the best approximation to a FP action I found.
It used the very expensive gauge action of Ref. \cite{INST}.
Action C has only the clover term but is otherwise identical.
My scaling tests of it used the new gauge action presented in
the Appendix.

I tested several other actions. All the variants of tadpole improvement
I studied had large mass renormalization. A pure hypercube action with
no Pauli term had a good dispersion relation at the coarsest lattice spacing,
but its hyperfine splittings were basically identical to those of
the Wilson action.

\subsection{Spectroscopy}

Lattice volumes were $4^3\times 16$ at $aT_c=1/2$ (excessively
 long in the time direction, in retrospect),
$6^3\times 16$ at $aT_c=1/3$, and $8^3\times 16$ (dangerously short) and
$8^3\times 24$ (safer) at $aT_c=1/4$.

The data set for Action A consists of 80 lattices at $aT_c=1/2$,
50 lattices at
 $aT_c=1/3$ and 36 $8^3\times 16$ lattices at $aT_c=1/4$.
The data set for Action C consists of 80 lattices at $aT_c=1/2$
and $aT_c=1/3$ and 60 $8^3\times 24$ lattices at $aT_c=1/4$.

The spectroscopy measurement is entirely straightforward. I gauge
fixed to Coulomb gauge and used a Gaussian independent particle source
wave function $\psi(r) =
\exp(-\gamma r^2)$ with $\gamma=1$, 0.5, and 0.25 at $aT_c=2,3,4$.
I used pointlike sinks projected onto low momentum states.
I used naive currents ($\bar \psi \gamma_5 \psi$, etc.)
 for interpolating fields.
The spectra appeared to be asymptotic (as shown by good (correlated)
 fits to a single exponential)
beginning at $t\simeq 2$ (at $aT_c=1/2$), 3-5 (at $aT_c=1/3$)
and 5-7 (at $aT_c=1/4)$ and the best fits were selected using the 
old HEMCGC criterion \cite{HEMCGC}.

My fiducial for comparison, simply because there are extensive data sets,
is Wilson-action quenched spectroscopy.  I have tried to restrict the data
I used for comparison to lattices with the proper physical volume.
I constructed my own $aT_c=1/2$ and $aT_c=1/3$ Wilson data sets
($\beta=5.1$ and 5.54)
since I could not find any results for these.
I also ran off 40 Wilson lattices at $\beta=5.7$ ($N_t=4)$ to measure
a dispersion relation. At that coupling my masses were within
a standard deviation of the much superior data set of Butler, et al.
\cite{BUTLER}.

To compare with the more standard improved actions, I also
performed a fiducial study using the clover action (Wilson fermion
action plus on-site clover term) in a background of Wilson action gauge fields.
  I tadpole-improved the action using 
$u_0 = ({\rm Tr} U_p /3)^{1/4}$ where $U_p$ is the
average plaquette. Data are at $aT_c=1/2$ and $aT_c=1/3$, where $u_0=0.802$
and 0.844.  This data set was 180 and 60 lattices at the two couplings.

\begin{figure}
\centerline{\ewxy{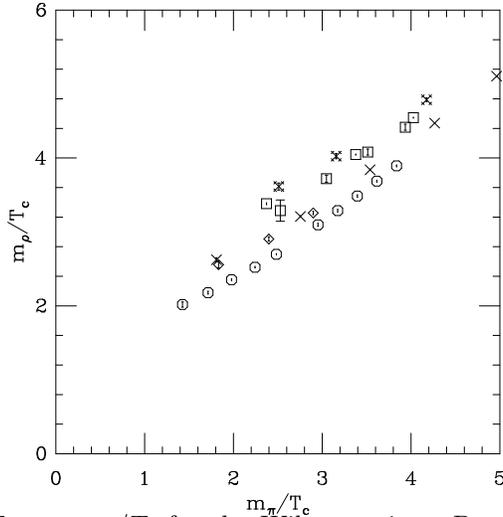}{80mm}
}
\caption{$m_\rho/T_c$ vs. $m_\pi/T_c$ for the Wilson action.
Data are labelled with 
octagons for $aT_c=1/2$,
diamonds for $aT_c=1/3$,
crosses for $aT_c=1/4$,
squares for $aT_c=1/8$,
and
fancy crosses for $aT_c=1,12$.
}
\label{fig:rhotcw}
\end{figure}

\begin{figure}
\centerline{\ewxy{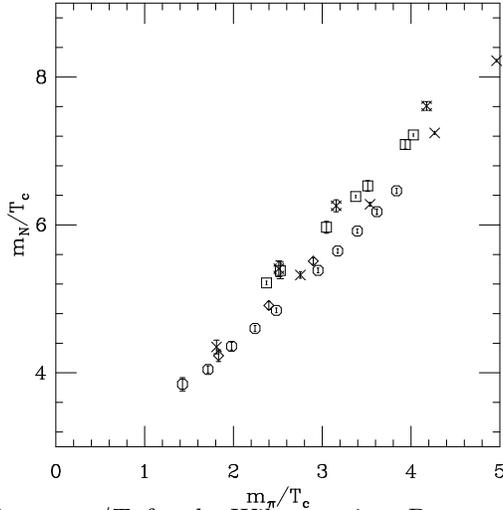}{80mm}
}
\caption{$m_N/T_c$ vs. $m_\pi/T_c$ for the Wilson action.
Data are labelled as in Fig. 11.}
\label{fig:ntcw}
\end{figure}

 I show first plots of
$m_\rho/T_c$ and $m_N/T_c$ vs. $m_\pi/T_c$ with Wilson fermions. 
These plots are scaling tests by themselves, or one can
interpolate in the curves to fixed values of $m_\pi/T_c$ (equivalent
to fixed quark mass) and plot the variation in the observable vs.
$aT_c$. Fig. \ref{fig:rhotcw}
 shows the rho mass
and Figs. \ref{fig:ntcw}
shows the nucleon mass.
 Notice that the rho mass
has the worst scaling violations of the three particles.

\begin{figure}
\centerline{\ewxy{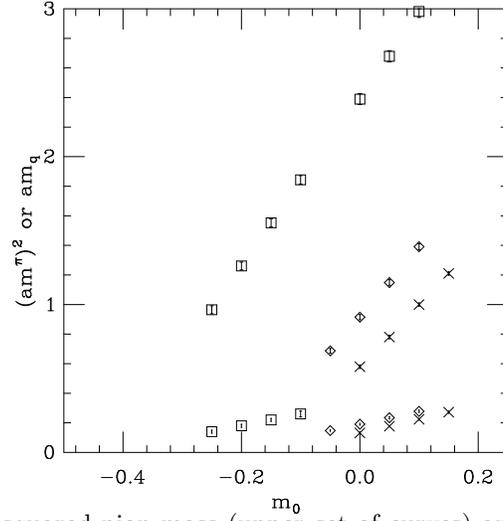}{80mm}
}
\caption{Bare squared pion mass  (upper set of curves) and
quark mass from Eq. 40 (lower set of curves) 
 vs bare quark mass for action A, at
$\beta_c$ for $N_t=2$ (squares), 3 (diamonds), 4 (crosses).}
\label{fig:pisqa}
\end{figure}

\begin{figure}
\centerline{\ewxy{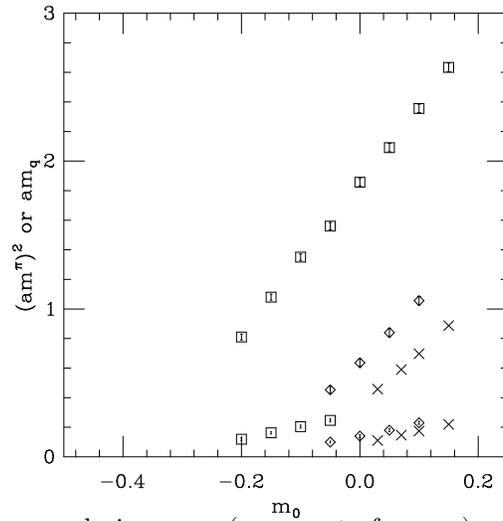}{80mm}
}
\caption{Bare squared pion mass (upper set of curves) and
quark mass from Eq. 40 (lower set of curves) vs bare quark mass for action C, at
$\beta_c$ for $N_t=2$ (squares), 3 (diamonds), 4 (crosses).}
\label{fig:pisqcl}
\end{figure}

We can roughly estimate the critical bare quark
mass (at which the pion is massless)
by  linearly extrapolating $m_\pi^2$ to zero in $m_0$.
Fig. \ref{fig:pisqa} shows the
squared pion mass vs bare quark mass for action A, at
$\beta_c$ for $N_t=2$, 3, 4, and the same plot, but for action C,
is shown in Fig. \ref{fig:pisqcl}. 
Both actions have small bare mass
renormalization.  
This is important from the point of view of principle
because a true FP action would have no additive mass renormalization.
It is important in practice because we only really know the
kinetic parameters by solving the RG equation
for positive bare mass; they must be extrapolated in
some artistic way if one needs to go to negative bare mass.

Another way to estimate the critical bare mass is to use
or  the PCAC relation
\bee
\nabla_\mu \cdot \langle \bar\psi \gamma_5 \psi(0) \bar\psi\gamma_5 \gamma_\mu
\psi(x) \rangle =
2m_q \langle\bar\psi \gamma_5 \psi(0) \bar\psi\gamma_5 \psi(x) \rangle .
\ee
If we  convert to lattice operators,
sum over spatial slices, and measure distance in the $t$ direction,
this becomes:
\bee
Z_A{\partial\over{\partial t}} \sum_{x,y,z} \langle \bar\psi \gamma_5 \psi(0)
\bar\psi\gamma_5 \gamma_0 \psi(x) \rangle =
2a m_q Z_P\sum_{x,y,z} \langle\bar\psi \gamma_5 \psi(0) \bar\psi\gamma_5 \psi(x) \rangle. 
\ee
I follow \cite{DOUGFPI} by fitting the pseudoscalar source-pseudoscalar sink to
\bee
P(t) = Z(\exp(-m_\pi t) + \exp(-m_\pi (N_t -t) ) )
\ee
and the pseudoscalar source-axial sink to
\bee
A(t) = {Z_P \over Z_A}
{{2m_q}\over m_\pi} Z(\exp(-(m_\pi t) - \exp(-m_\pi (N_t -t) )).
\ee
to extract $m_q$.
There are many other possibilities for defining an axial current and for
defining the derivative operator.
I only use the naive (pointlike) currents.  I do not know the
$Z-$ factors, but for finding the value of $m_0^c$ that does not matter.
Extrapolating $m_\pi^2$ or $m_q$ linearly in $m_0$
 ignores all the well-known  problems associated with
extracting quark masses from lattice data \cite{GUPTA},
 but the procedure is perfectly adequate to
distinguish a small quark mass from a large one.
The quark masses are shown in Figs. \ref{fig:pisqa} and \ref{fig:pisqcl}.

We find for action A that $m_0^c= -0.42$, -0.20, and -0.14 at
$aT_c=1/2$, 1/3, and 1/4, respectively.
For action C, the corresponding numbers are
$m_0^c= -0.36$, $-0.16$, and $-0.095$. 
These numbers should be compared to the analogous quantities for Wilson fermions, using $m_0^c = 1/(2\kappa_c) -4$: -1.58 at $\beta=5.1$ ($aT_c=1/2$),
-1.04 at $\beta=5.7$ ($aT_c=4$), and still -0.70 at $\beta=6.3$ \cite{LM}.
In the latter case tadpole improved perturbation theory can explain
most of the mass shift.

\begin{figure}
\centerline{\ewxy{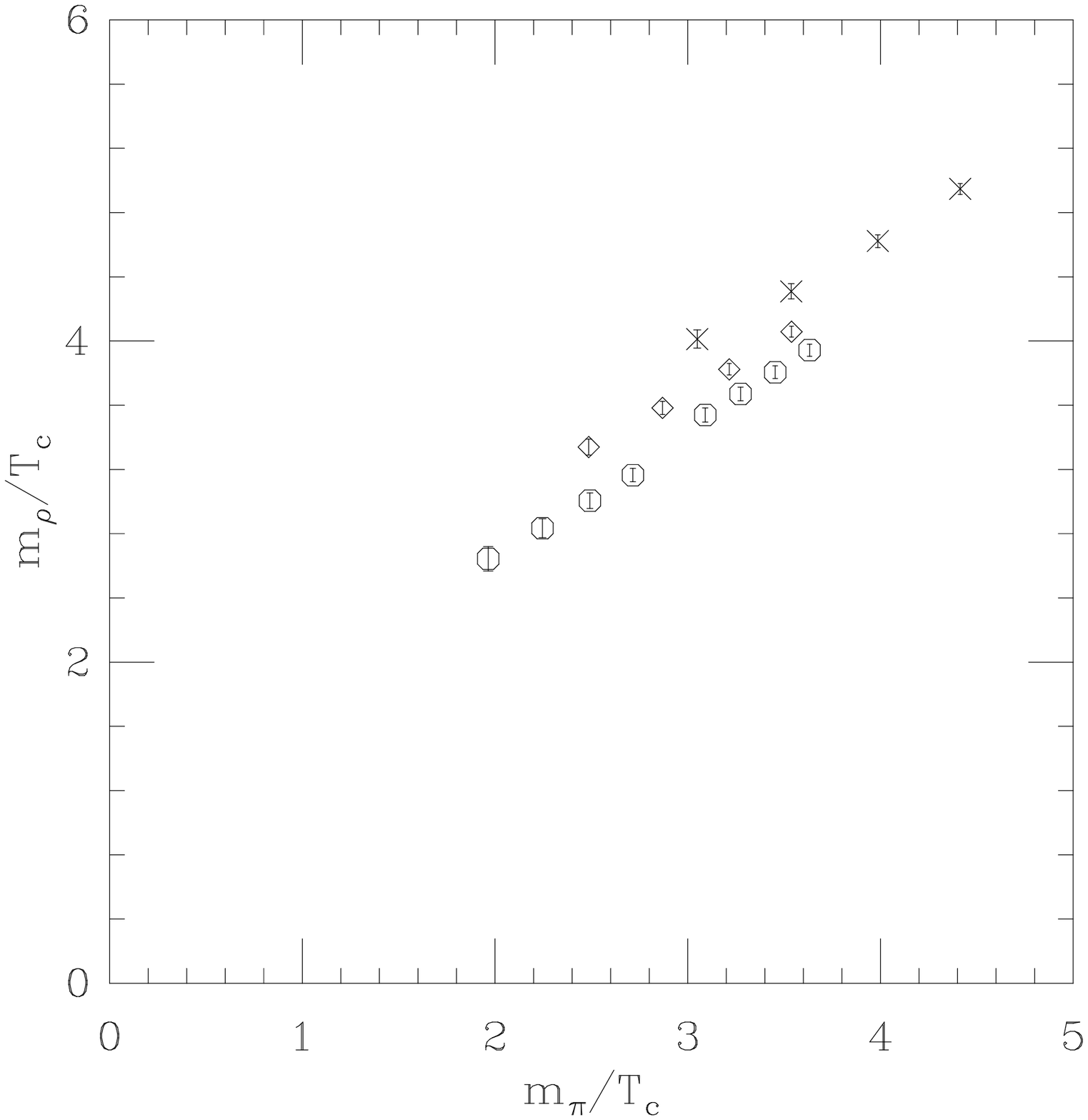}{80mm}
}
\caption{$m_\rho/T_c$ vs. $m_\pi/T_c$ for  action A.}
\label{fig:rhotca}
\end{figure}

\begin{figure}
\centerline{\ewxy{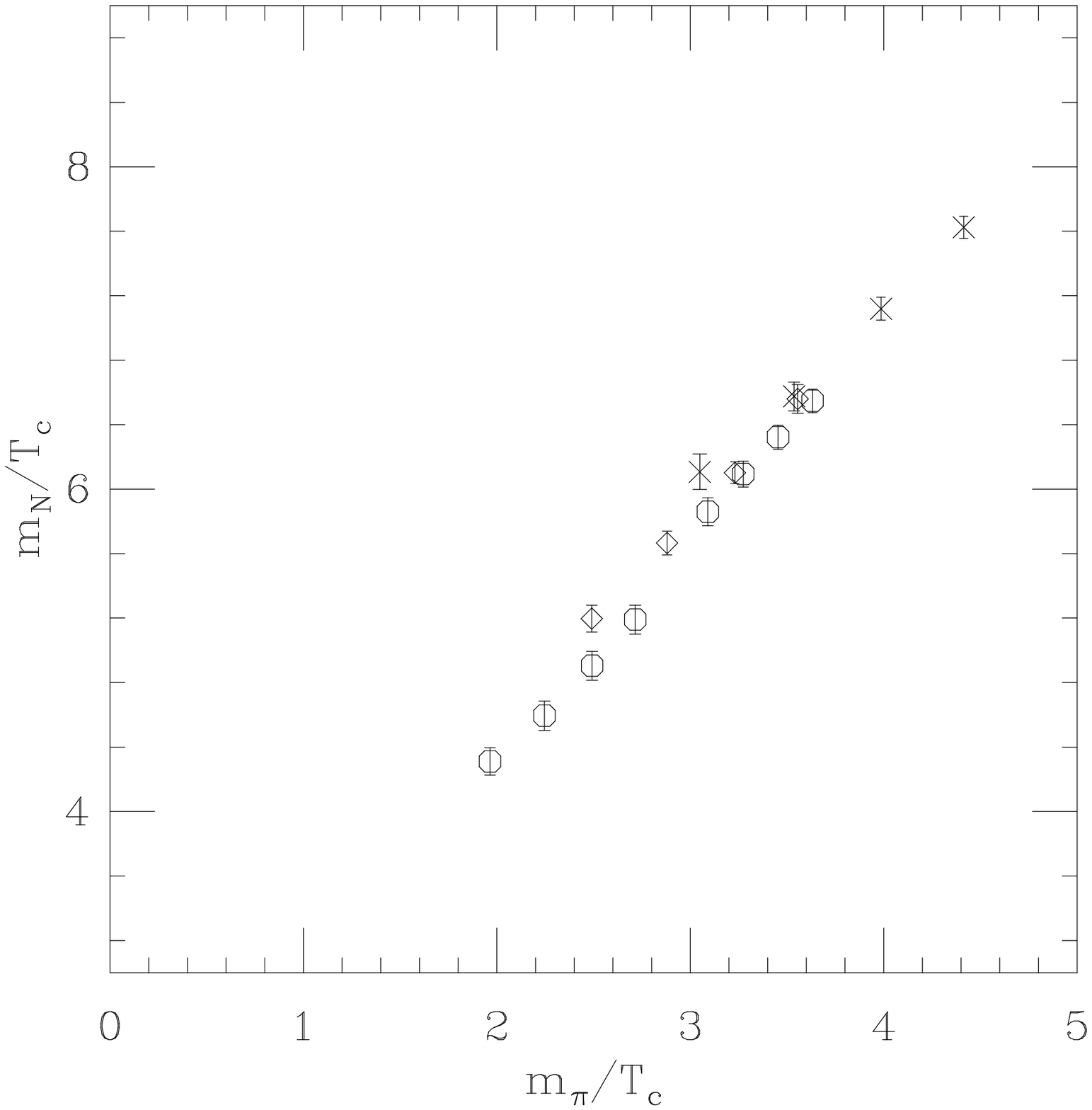}{80mm}
}
\caption{$m_N/T_c$ vs. $m_\pi/T_c$ for  action A.}
\label{fig:ntca}
\end{figure}

\begin{figure}
\centerline{\ewxy{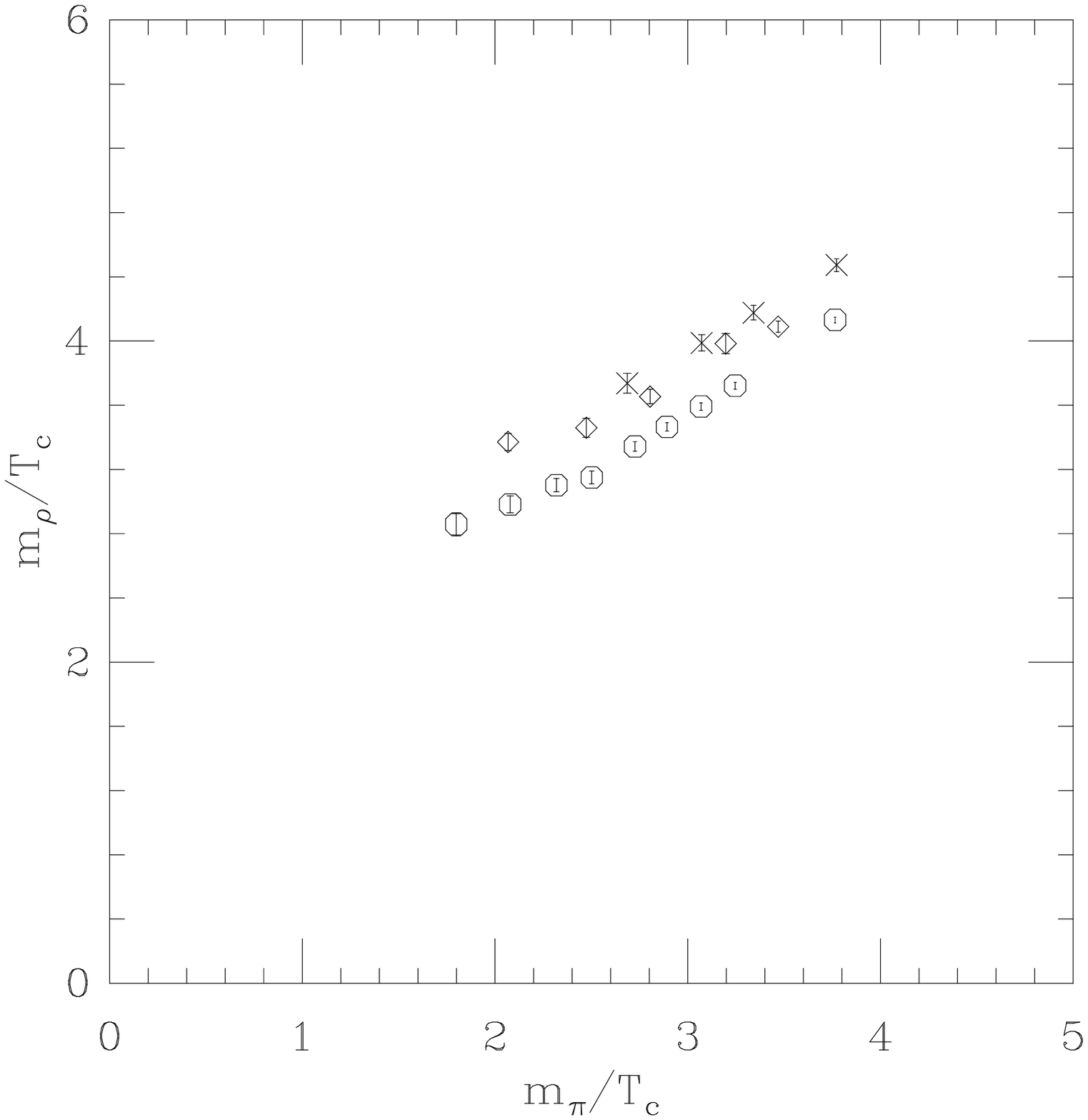}{80mm}
}
\caption{$m_\rho/T_c$ vs. $m_\pi/T_c$ for  action C.}
\label{fig:rhotccl}
\end{figure}

\begin{figure}
\centerline{\ewxy{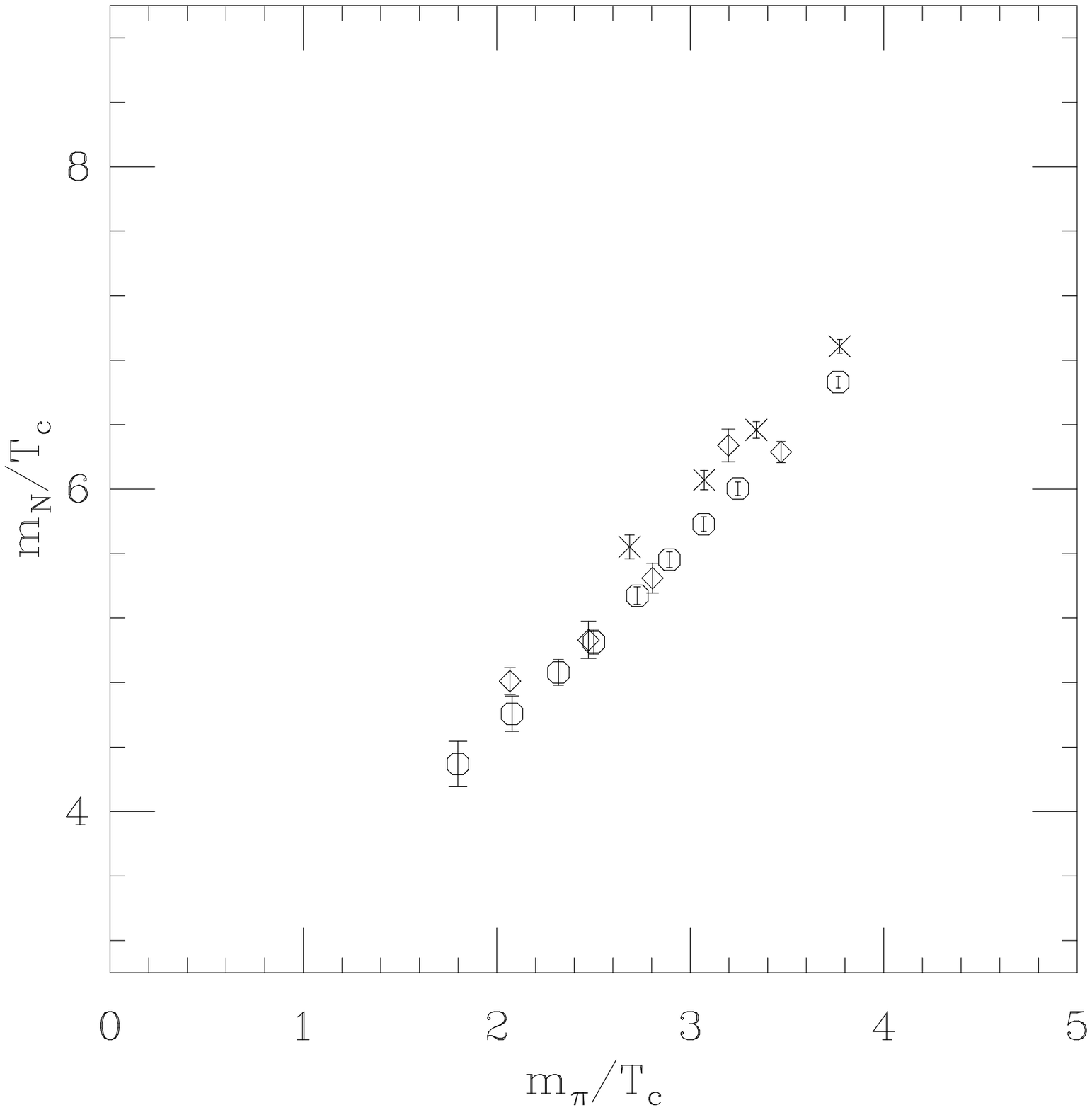}{80mm}
}
\caption{$m_N/T_c$ vs. $m_\pi/T_c$ for  action C.}
\label{fig:ntccl}
\end{figure}

Now for scaling tests. I compare
$m_\rho/T_c$ and $m_N/T_c$ vs. $m_\pi/T_c$ for actions A and C in
 Figs. \ref{fig:rhotca},  \ref{fig:ntca},  \ref{fig:rhotccl},
and  \ref{fig:ntccl}.

\begin{figure}
\centerline{\ewxy{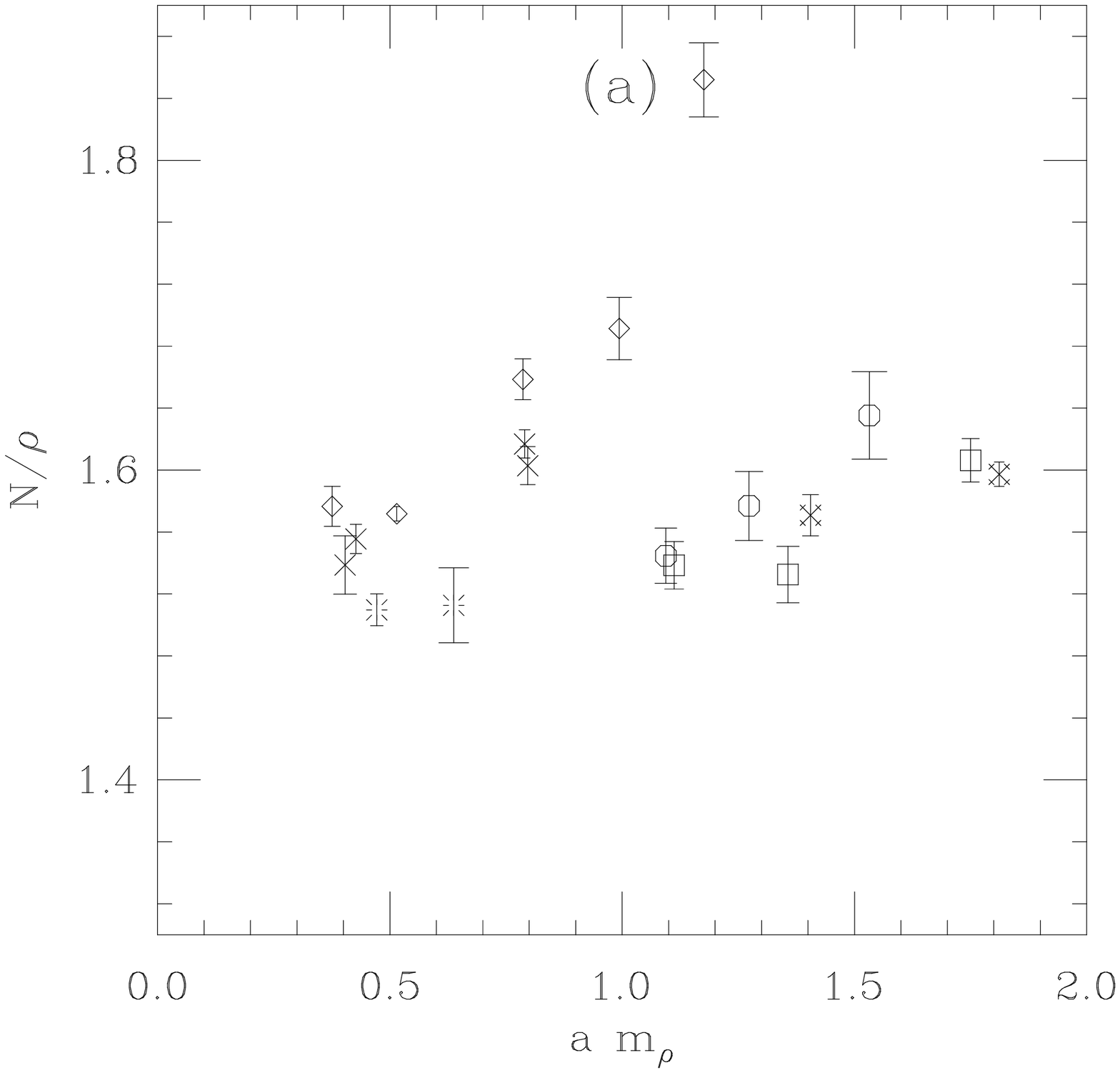}{80mm}
\ewxy{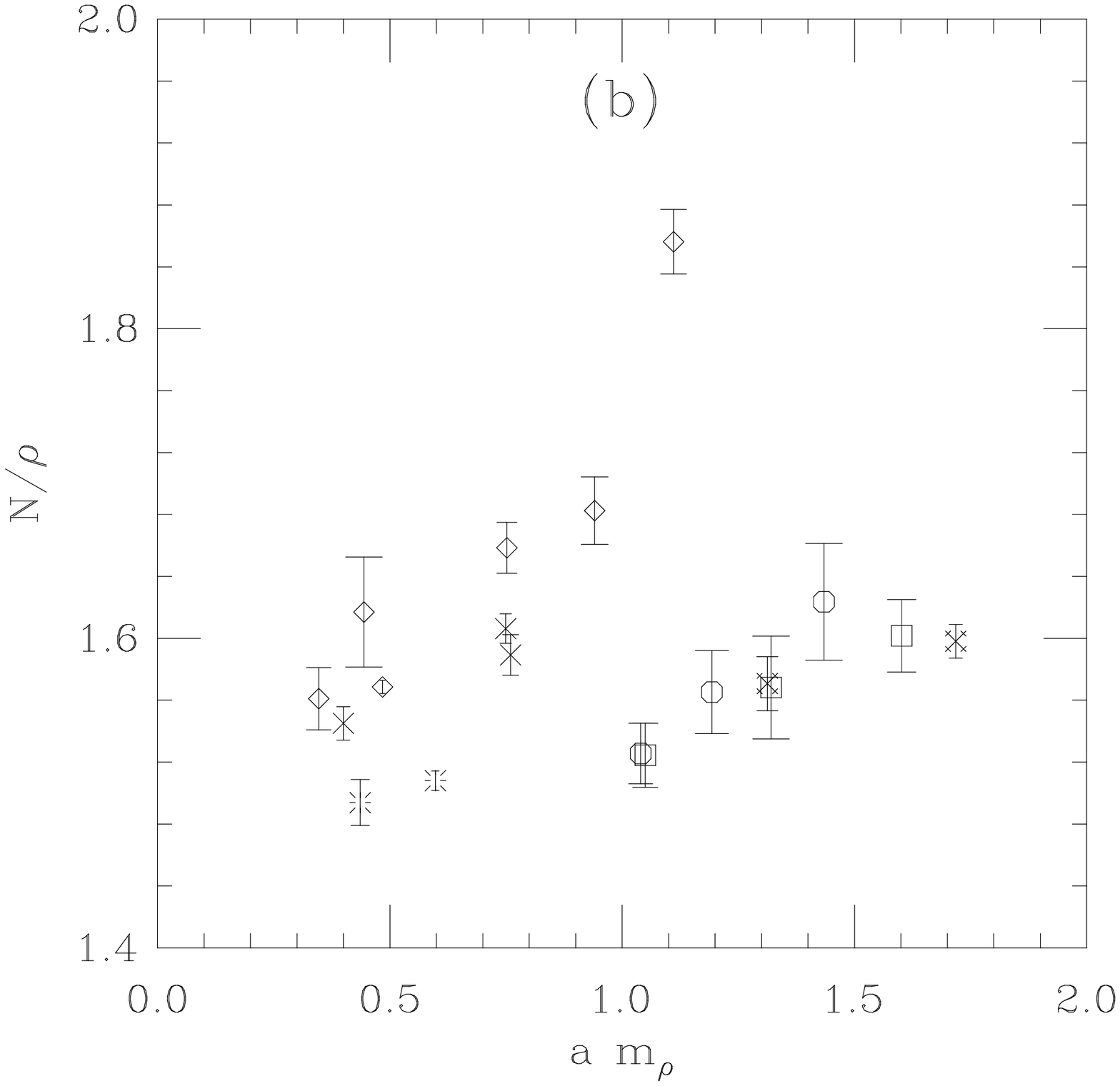}{80mm}}
\vspace{0.5cm}
\centerline{\ewxy{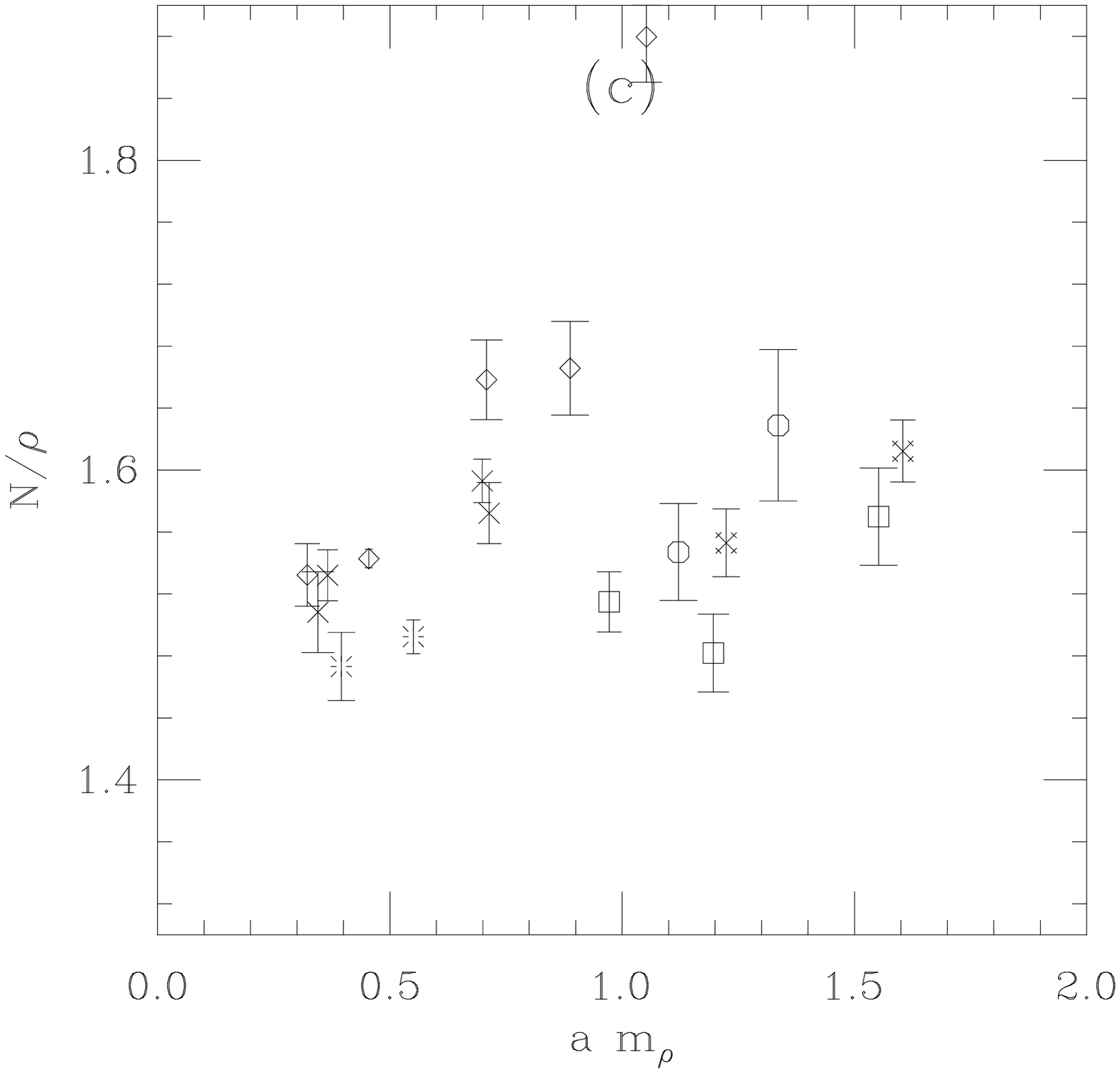}{80mm}
\ewxy{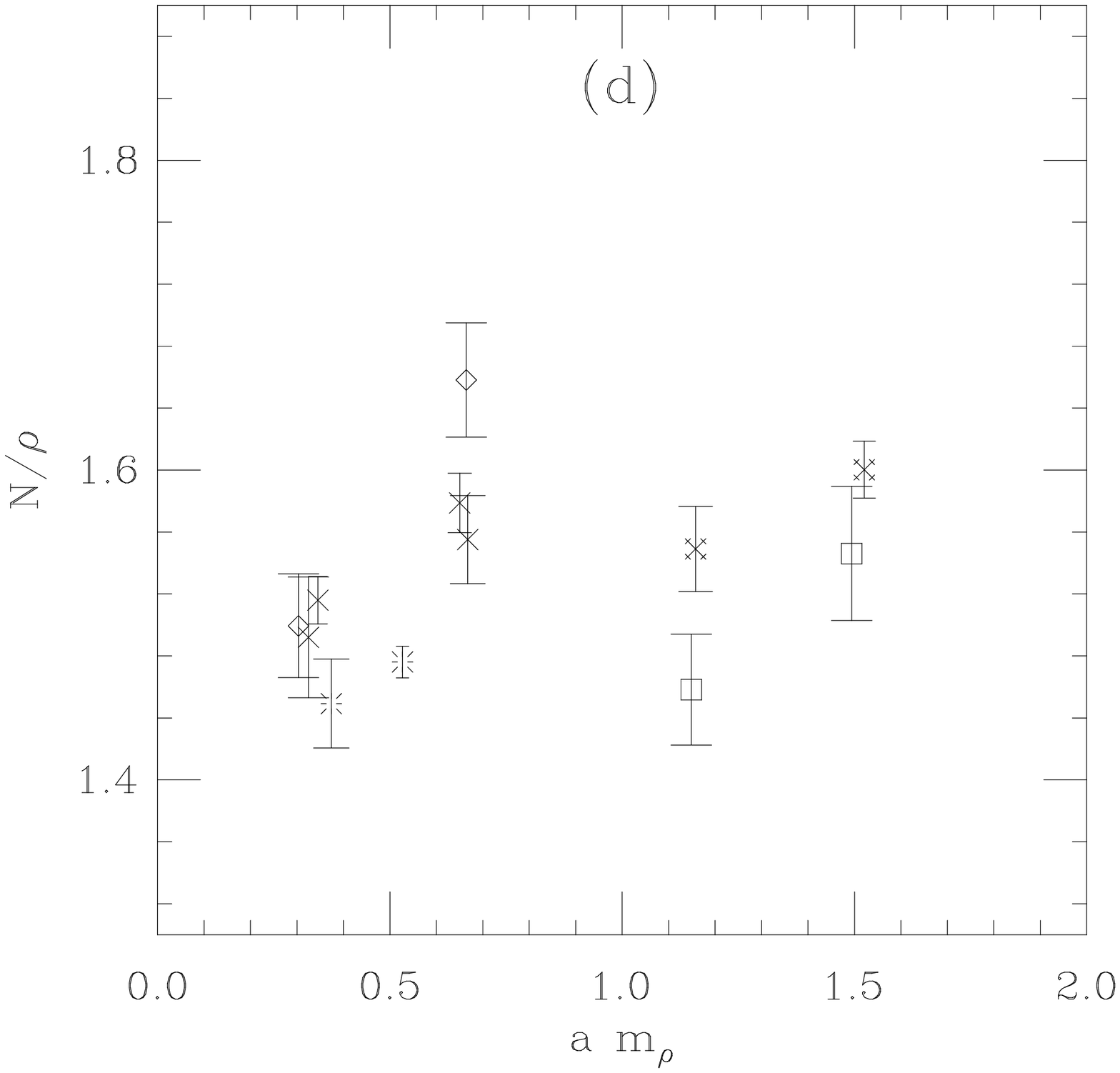}{80mm}}
\caption{A scaling test for new actions: (squares action C,
 octagons action A)
vs. Wilson actions on lattices of fixed physical size (diamonds)
and larger volumes (crosses), and the
nonperturbatively improved  (bursts) and tadpole improved (fancy crosses)
clover actions.
 Data are interpolated to
$\pi/\rho=0.84$ (a), 0.80 (b), 0.75 (c) and 0.70 (d).}
\label{fig:allrat}
\end{figure}

We can compare scaling violations in hyperfine splittings by
interpolating our data to fixed $\pi/\rho$ mass ratios and plotting
the $N/\rho$ mass ratio vs. $m_\rho a$. I do this at four
$\pi/\rho$ mass ratios, 0.80 and 0.70, in Fig. \ref{fig:allrat}.
In these figure the diamonds are Wilson action data in lattices of fixed
physical size
 ($4^3$ at $\beta=5.1$,
$6^3$ at $\beta=5.54$,
$8^3$ at $\beta=5.7$ \cite{BUTLER},
$16^3$ at $\beta=6.0$ \cite{DESY96173}
$24^3$ at $\beta=6.3$ \cite{APE63})
 and the crosses are data in various larger lattices:
$16^3$ and $24^3$ at $\beta=5.7$ and $32^3$ at $\beta=6.17$ \cite{BUTLER},
$24^3$ at $\beta=6.0$ \cite{DESY96173}.
When they are present the data points from larger lattices illustrate
the danger of performing scaling tests with data from different volumes.
The bursts are from the nonperturbatively improved clover action of Ref.
\cite{ALPHA} and the fancy diamonds are the TI clover action.
The other plotting symbols show our test actions A and C.

We give a tables of masses from Action C in Tables
\ref{tab:m3092}, 
\ref{tab:m350}, 
and \ref{tab:m370},
and for Action A in Tables
\ref{tab:m238}, 
\ref{tab:m285}, 
and \ref{tab:m305}.

\subsection{Dispersion Relations}

\begin{figure}[htb]
\begin{center}
\vskip -10mm
\leavevmode
\epsfxsize=60mm
\epsfbox[40 50 530 590]{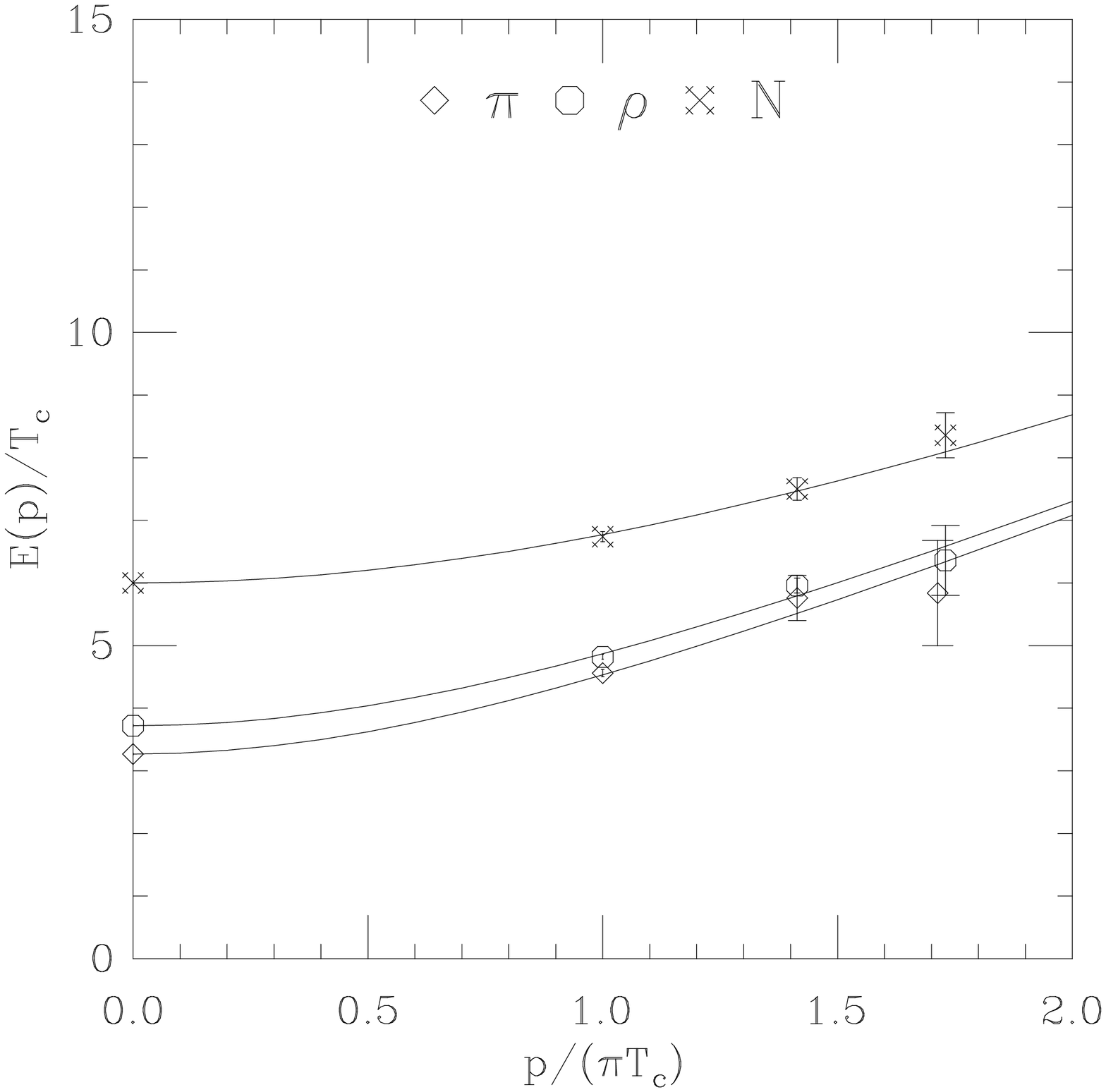}
\vskip -5mm
\end{center}
\caption{Dispersion relation for heavy hadrons at $aT_c=1/2$ ($a \simeq 0.36$
fm)  from action C.
The curves are the continuum dispersion relation for the appropriate
(measured) hadron mass.}
\label{fig:eptc}
\end{figure}

There are two ways to look at dispersion relations. The simplest is to
plot $E(p)$ the energy of the state produced with spatial momentum
$\vec p$, as a function of $|\vec p|$.
The result for   action C  at bare mass 0.15 is compared to the
free dispersion relation 
at $aT_c=1/2$ in Fig. \ref{fig:eptc}.
All of the test actions I have studied have good dispersion relations
even at $aT_c=1/2$. I believe that is a generic feature of the
hypercubic kinetic term.

\begin{figure}
\centerline{\ewxy{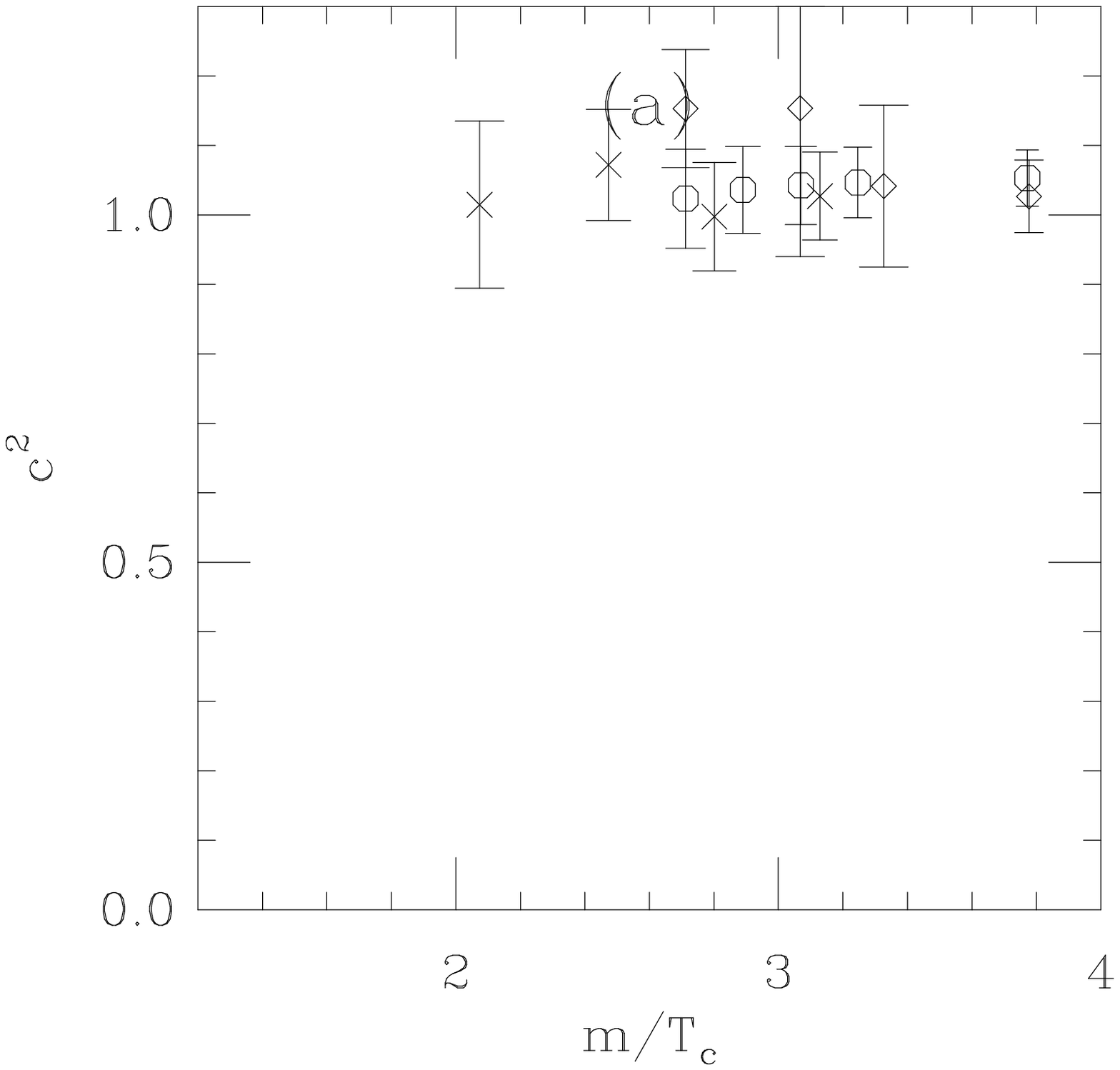}{80mm}
\ewxy{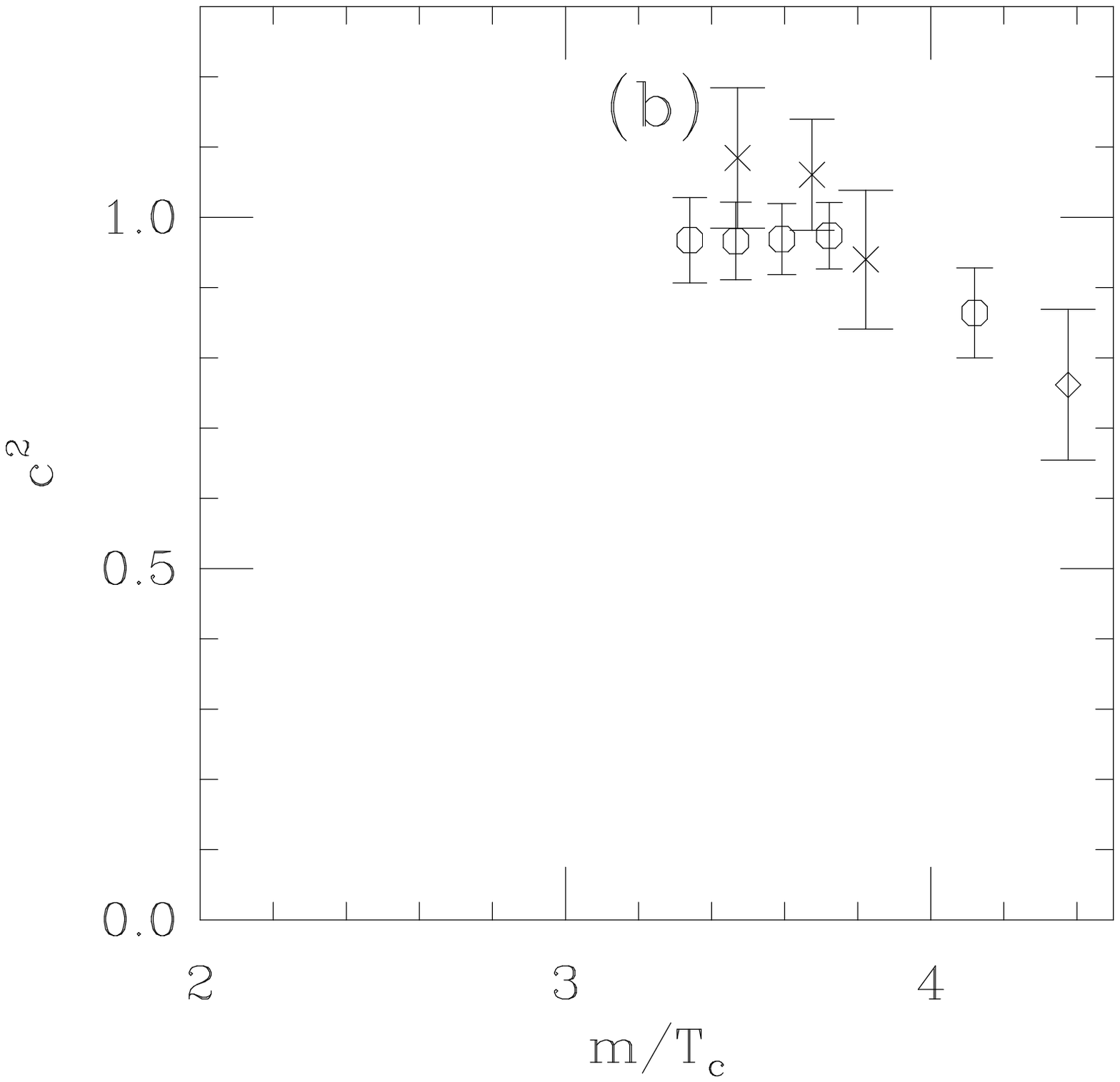}{80mm}}
\vspace{0.5cm}
\centerline{\ewxy{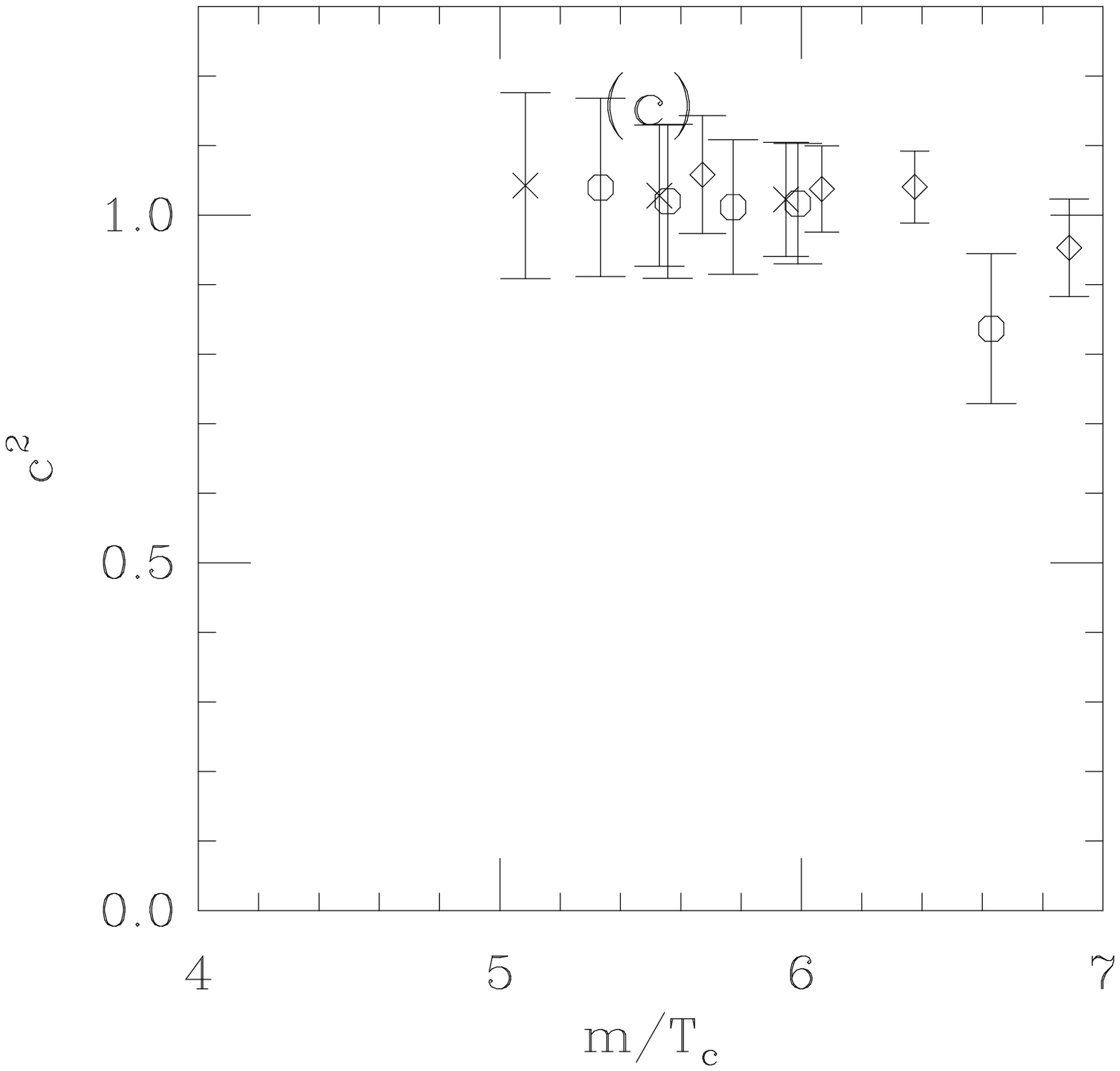}{80mm}
}
\caption{ Squared speed of light vs. hadron mass in units of $T_c$,
for (a) pseudoscalars, (b) vectors) and (c) protons, from action C.
Octagons, crosses, and diamonds label $aT_c= 1/2$, 1/3 and 1/4.}
\label{fig:drcl}
\end{figure}

\begin{figure}
\centerline{\ewxy{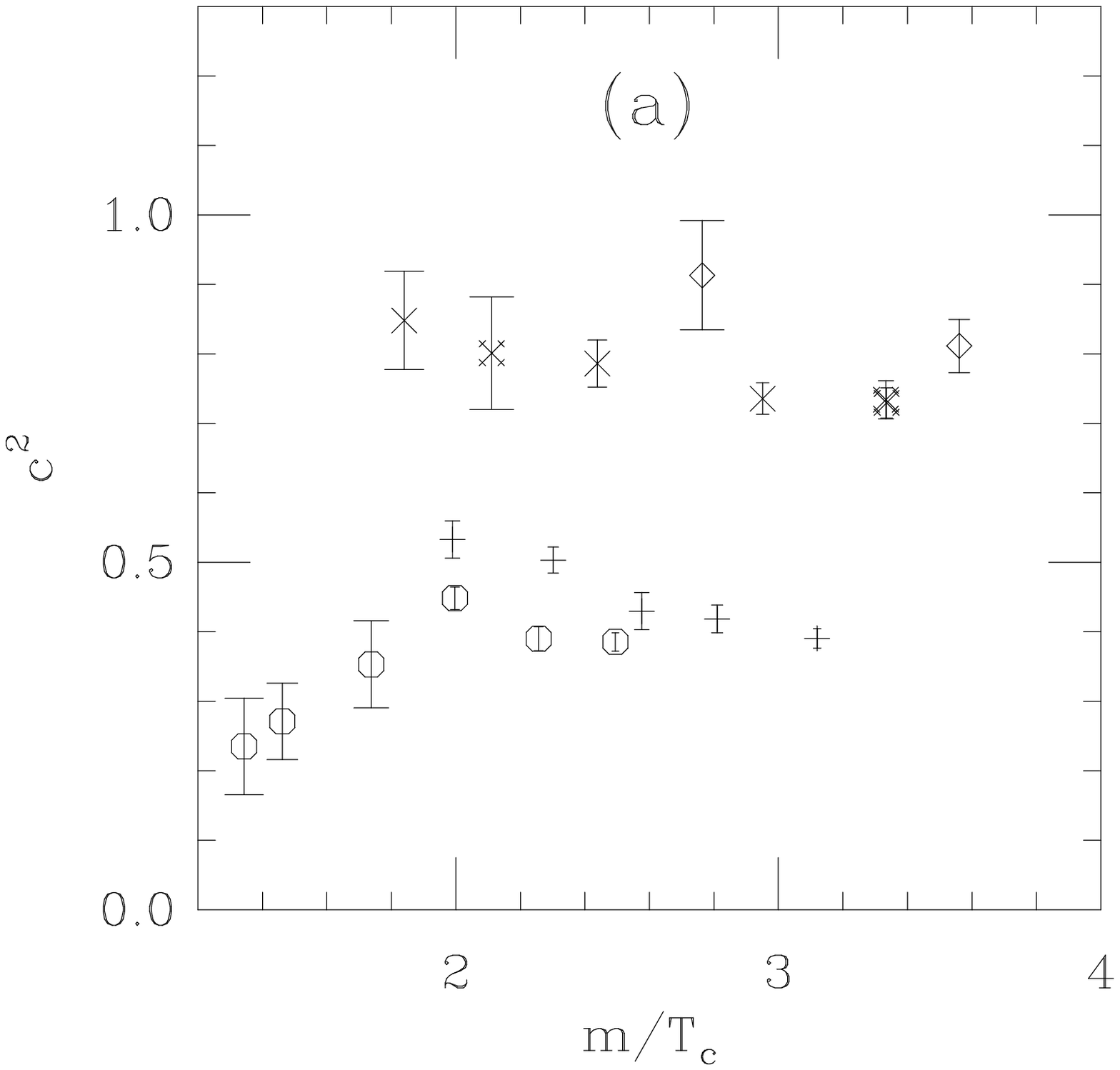}{80mm}
\ewxy{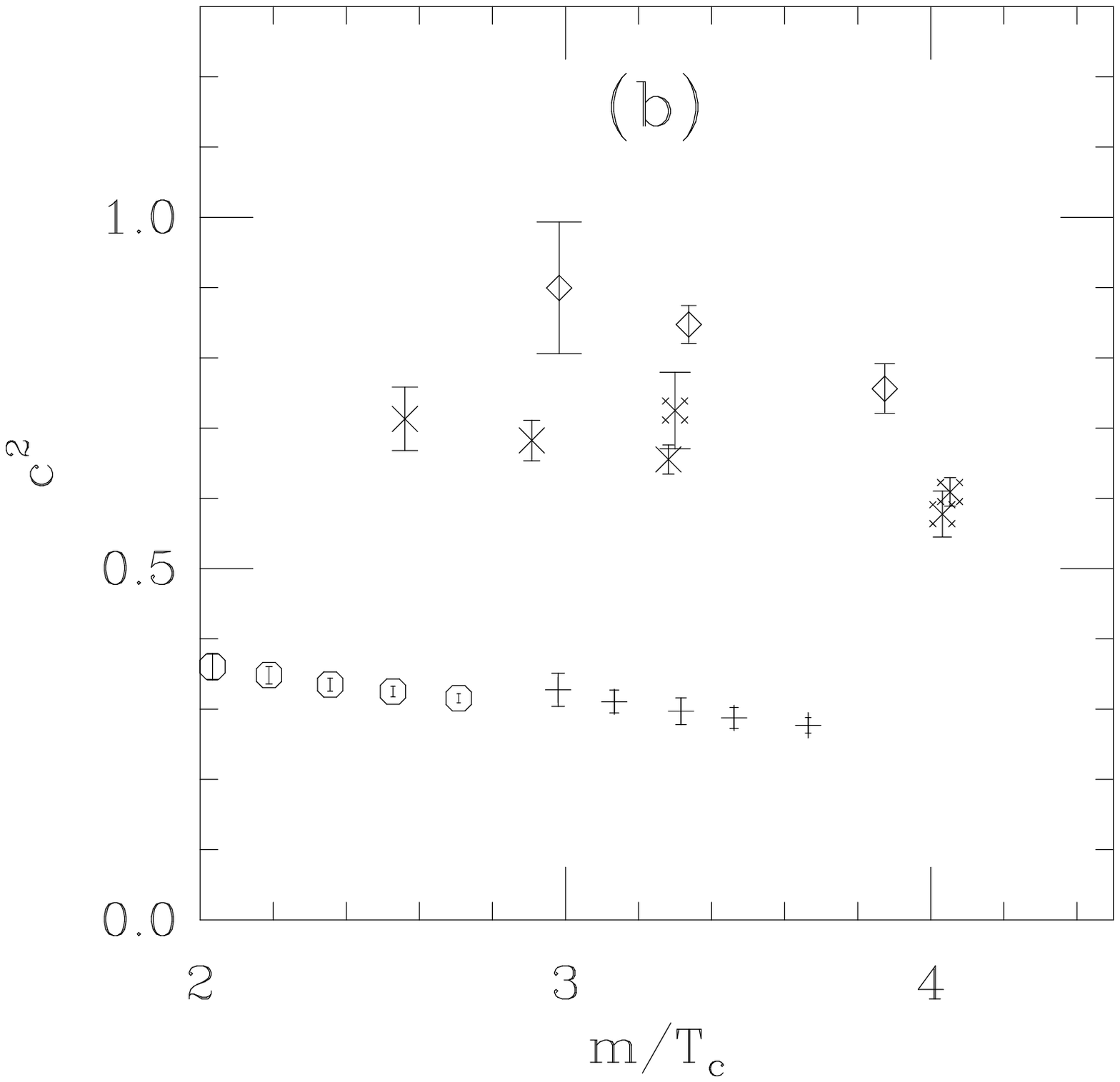}{80mm}}
\vspace{0.5cm}
\centerline{\ewxy{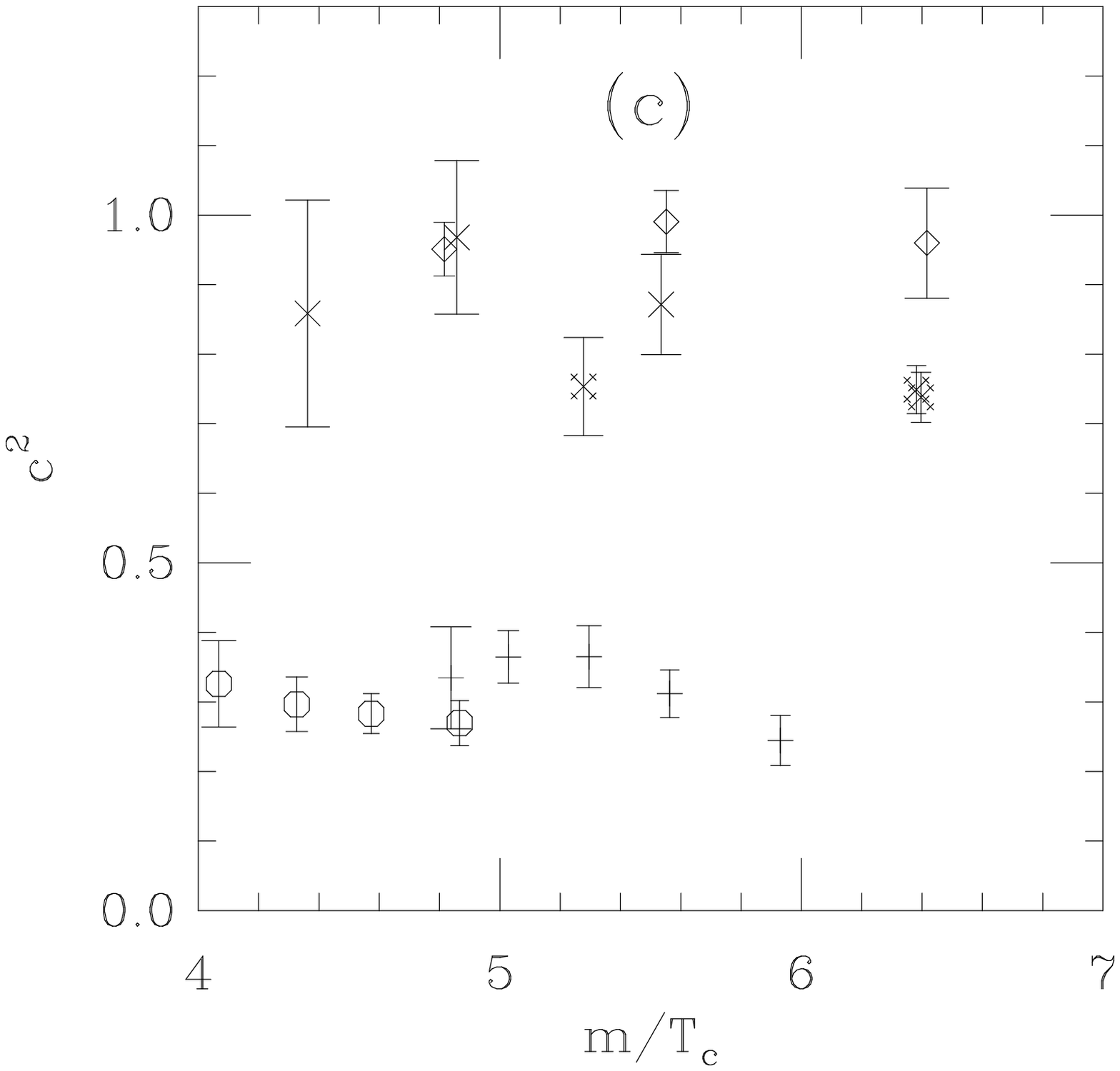}{80mm}
}
\caption{ Squared speed of light vs hadron mass in units of $T_c$,
for (a) pseudoscalars, (b) vectors) and (c) protons, from the Wilson and clover
actions.
Octagons, crosses, and diamonds label $aT_c= 1/2$, 1/3 and 1/4 for the 
Wilson action, and for the clover action the  labels are plusses for $aT_c=1/4$
and fancy crosses for $aT_c=1/3$.}
\label{fig:drw}
\end{figure}

The signal to noise ratio for the nonzero momentum meson channels
dies away at large $t$ like $\exp(-(E(p)-m_\pi)t)$.  This means that
large statistics are required to go to small quark mass or to high
$\vec p$.  However, it is possible to extract a fitted
$c^2_{eff}= (E(p)^2-m^2)/p^2 $ for the lowest nonzero momentum mode,
for larger masses.  This was done by performing a 4-parameter
correlated fit to a pair of single exponentials, one for the $\vec p=(0,0,0)$
mode and the other the $\vec p = (1,0,0)$ mode.
I compare my results from action C and from the Wilson  and clover actions in 
Figs. \ref{fig:drcl} and \ref{fig:drw}. Hadron masses are again scaled
by $T_c$ to allow the display of several lattice spacings at once.
Action C has $c^2_{eff} \simeq 1$ for all observed hadrons even at
$p = \pi T_c$ (860 MeV/c) at $aT_c=1/2$.

\begin{table}
\begin{tabular}{|c|l|l|l|l|l|}
\hline
$am_q$ & PS  & V  &  N  &  $\Delta$ \\
\hline
   0.30 &  1.883( 8) &  2.066( 8) &  3.332(17) &  3.450(21)  \\
   0.15 &  1.623( 9) &  1.861(10) &  3.002(20) &  3.156(25)  \\
   0.10 &  1.535( 9) &  1.796(11) &  2.891(22) &  3.061(27)  \\
   0.05 &  1.446( 9) &  1.733(12) &  2.781(24) &  2.967(30)  \\
   0.00 &  1.363( 9) &  1.672(14) &  2.670(27) &  2.875(33)  \\
   -0.05 &  1.251(10) &  1.575(20) &  2.526(37) &  2.770(46)  \\
   -0.10 &  1.159(10) &  1.551(20) &  2.432(39) &  2.687(43)  \\
   -0.15 &  1.039(12) &  1.491(26) &  2.303(54) &  2.591(49)  \\
   -0.20 &  0.899(16) &  1.429(35) &  2.147(70) &  2.505(64)  \\
\hline
\end{tabular}
\caption{Table of best-fit masses, action C, $\beta=3.092$ ($aT_c=1/2)$.}
\label{tab:m3092}
\end{table}

\begin{table}
\begin{tabular}{|c|l|l|l|l|l|}
\hline
$am_q$ & PS  & V  &  N  &  $\Delta$ \\
\hline
   0.15 &  1.156( 7) &  1.363(11) &  2.077(21) &  2.258(28)  \\
   0.10 &  1.066(15) &  1.328(20) &  2.090(33) &  2.277(51)  \\
   0.05 &  0.935( 8) &  1.218(15) &  1.816(30) &  2.049(31)  \\
   0.00 &  0.825( 9) &  1.153(19) &  1.688(38) &  1.959(41)  \\
   -0.05 &  0.689(10) &  1.124(18) &  1.603(27) &  1.941(33)  \\
\hline
\end{tabular}
\caption{Table of best-fit masses, action C, $\beta=3.50$ ($aT_c=1/3)$.}
\label{tab:m350}
\end{table}

\begin{table}
\begin{tabular}{|c|l|l|l|l|l|}
\hline
$am_q$ & PS  & V  &  N  &  $\Delta$ \\
\hline
   0.15 &  0.943( 4) &  1.118( 9) &  1.722(10) &  1.853(14)  \\
   0.10 &  0.836( 4) &  1.044(11) &  1.592(12) &  1.742(17)  \\
   0.07 &  0.768( 5) &  0.997(12) &  1.514(14) &  1.678(19)  \\
   0.03 &  0.672( 5) &  0.934(15) &  1.410(18) &  1.598(22)  \\
\hline
\end{tabular}
\caption{Table of best-fit masses, action C, $\beta=3.70$ ($aT_c=1/4)$.}
\label{tab:m370}
\end{table}

\begin{table}
\begin{tabular}{|c|l|l|l|l|l|}
\hline
$am_q$ & PS  & V  &  N  &  $\Delta$ \\
\hline
   0.15 &  1.816(11) &  1.972(18)&  3.274(36) &  3.355(44)  \\
   0.10 &  1.727(11) &  1.903(19)&  3.161(37) &  3.309(49)  \\
   0.05 &  1.637(11) &  1.835(21)&  3.047(39) &  3.237(47)  \\
   0.00 &  1.546(11) &  1.769(22)&  2.930(43) &  3.156(51)  \\
   -0.10 &  1.358(10) &  1.582(20)&  2.596(44) &  2.825(43)  \\
   -0.15 &  1.246(10) &  1.502(24)&  2.452(44) &  2.718(49)  \\
   -0.20 &  1.123(11) &  1.417(29)&  2.297(45) &  2.605(57)  \\
   -0.25 &  0.982(13) &  1.322(37)&  2.155(42) &  2.483(71)  \\
\hline
\end{tabular}
\caption{Table of best-fit masses, action A, $\beta=2.38$ ($aT_c=1/2)$.}
\label{tab:m238}
\end{table}

\begin{table}
\begin{tabular}{|c|l|l|l|l|l|}
\hline
$am_q$ & PS  & V  &  N  &  $\Delta$ \\
\hline
   0.10 &  1.179( 8) &  1.353(11)&  2.145(19) &  2.245(24)  \\
   0.05 &  1.072( 8) &  1.275(12)&  2.010(21) &  2.115(27)  \\
   0.00 &  0.956( 9) &  1.194(13)&  1.870(24) &  1.989(32)  \\
   -0.05 &  0.829(10) &  1.113(16)&  1.720(28) &  1.854(40)  \\
\hline
\end{tabular}
\caption{Table of best-fit masses, action A, $\beta=2.85$ ($aT_c=1/3)$.}
\label{tab:m285}
\end{table}

\begin{table}
\begin{tabular}{|c|l|l|l|l|l|}
\hline
$am_q$ & PS  & V  &  N  &  $\Delta$ \\
\hline
   0.15 &  1.100( 5) &  1.228( 8)&  1.903(20) &  1.998(27)  \\
   0.10 &  0.993( 6) &  1.146( 9)&  1.783(18) &  1.875(30)  \\
   0.05 &  0.880( 6) &  1.066(11)&  1.640(20) &  1.788(26)  \\
   0.00 &  0.758( 7) &  1.000(12)&  1.504(25) &  1.662(29)  \\
\hline
\end{tabular}
\caption{Table of best-fit masses, action A, $\beta=3.05$ ($aT_c=1/4)$.}
\label{tab:m305}
\end{table}

\subsection{Summary}

It appears that these actions are members of  a  family of actions which show
improved scaling, even at $\beta_c(N_t=2)$, about 0.36 fm lattice spacing.

The hypercubic actions have much better dispersion relations than
either the clover or Wilson action.  They share this improvement
with the D234 family of actions \cite{ALFORD} and with the 
Hamber-Wu \cite{HW} action as tested in Ref. \cite{LEIN}.

However, the hypercubic actions tested here
seem to produce about the same level of improvement in hyperfine
splittings as the clover action, at heavier quark masses.
Leaving out the Pauli term gives noticeable scaling violations
with a too-large $N/\rho$ ratio; probably one needs to keep some kind of
explicit Pauli/clover term in the lattice action
to boost the hyperfine splittings. 

 The hyperfine splittings 
show worse scaling violations than the dispersion relation. 
Controlling and approximating the quark anomalous magnetic moment
is the most difficult part of the construction of a FP action, and
that may be the source of the scale violations.

Of course, there is still the possibility that all the actions tested here
are missing some other common physics ingredient, which is responsible
for scaling of the hyperfine splittings.

\begin{figure}
\centerline{\ewxy{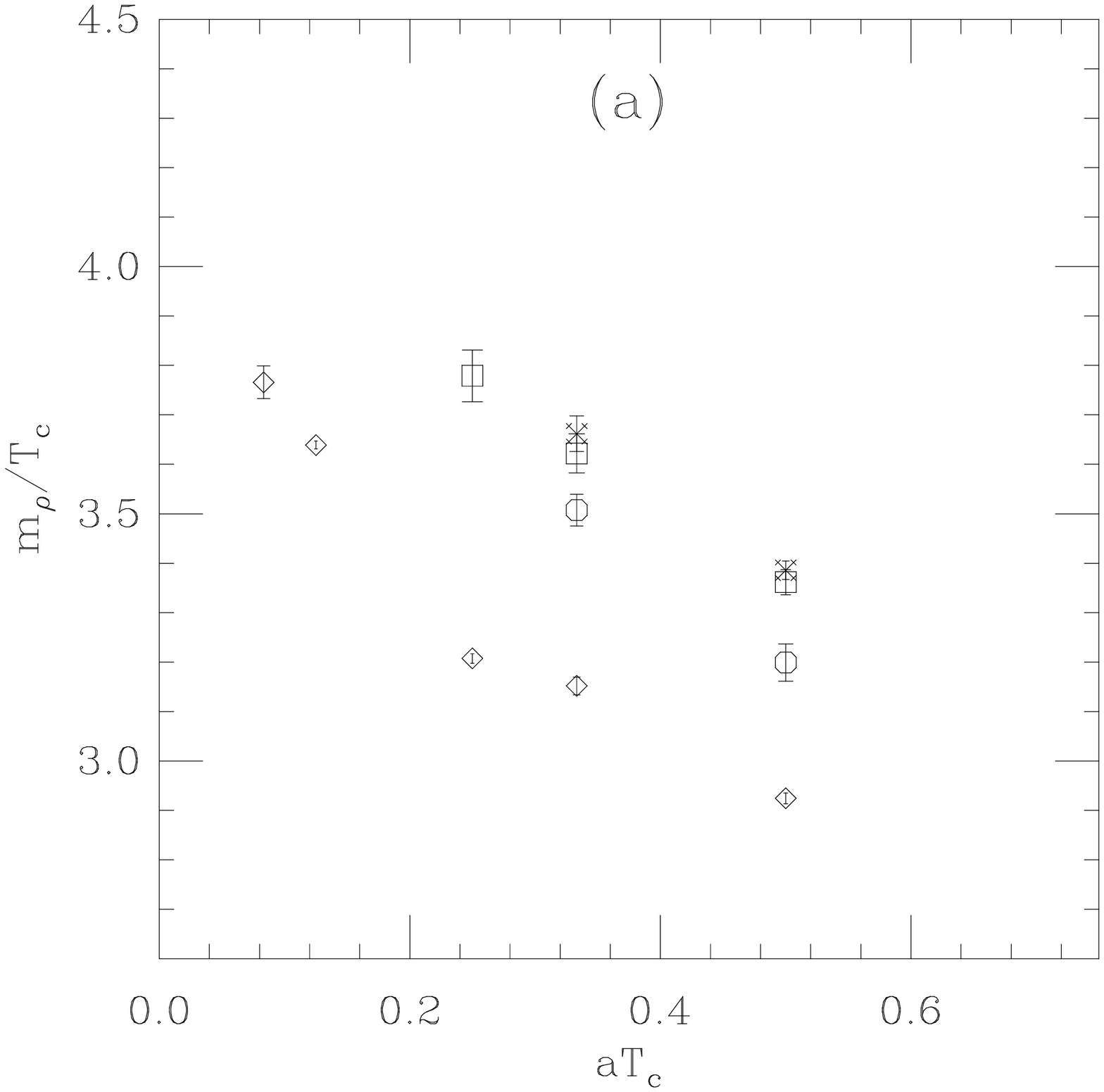}{80mm}
\ewxy{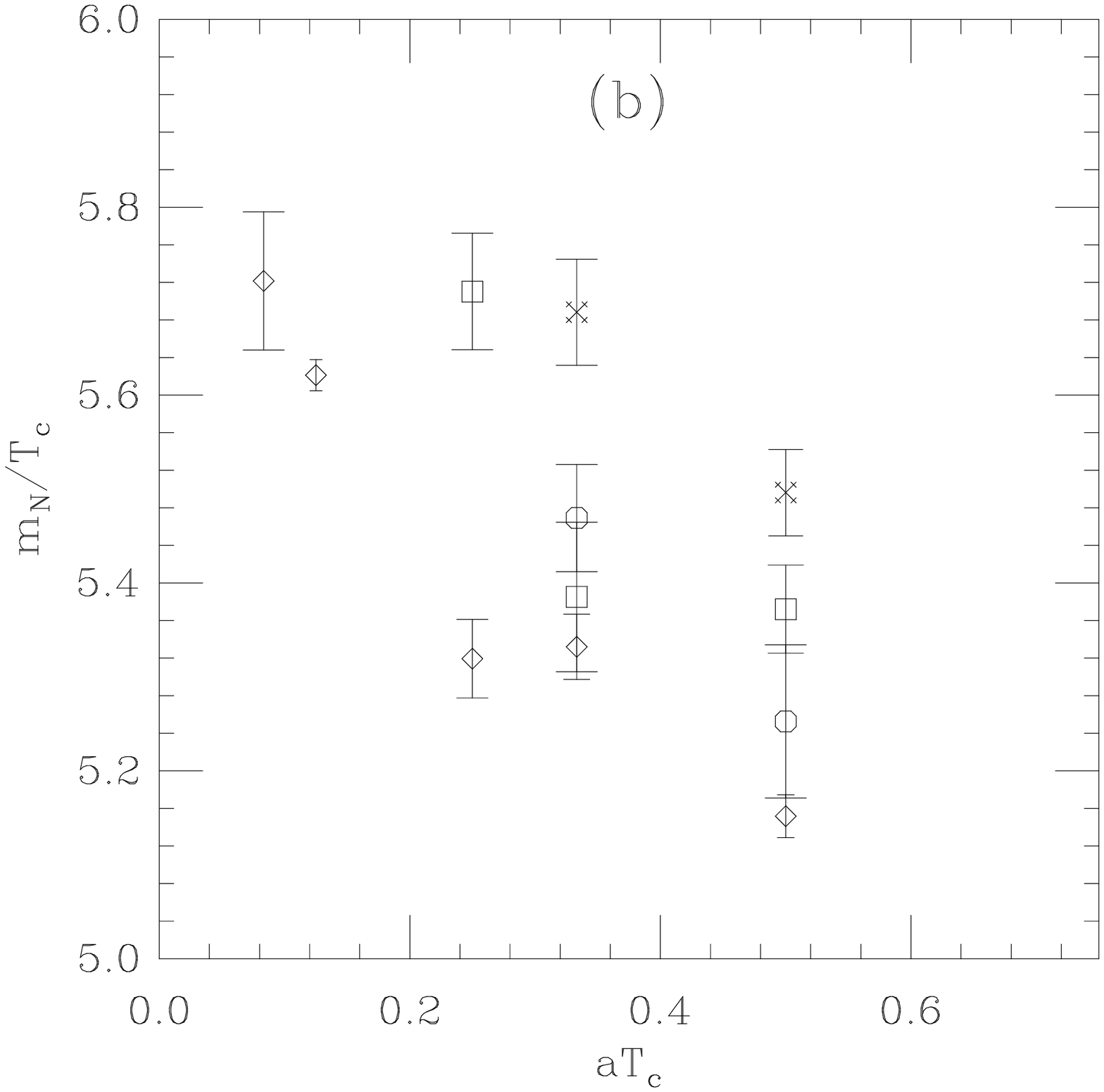}{80mm}}
\vspace{0.5cm}
\centerline{\ewxy{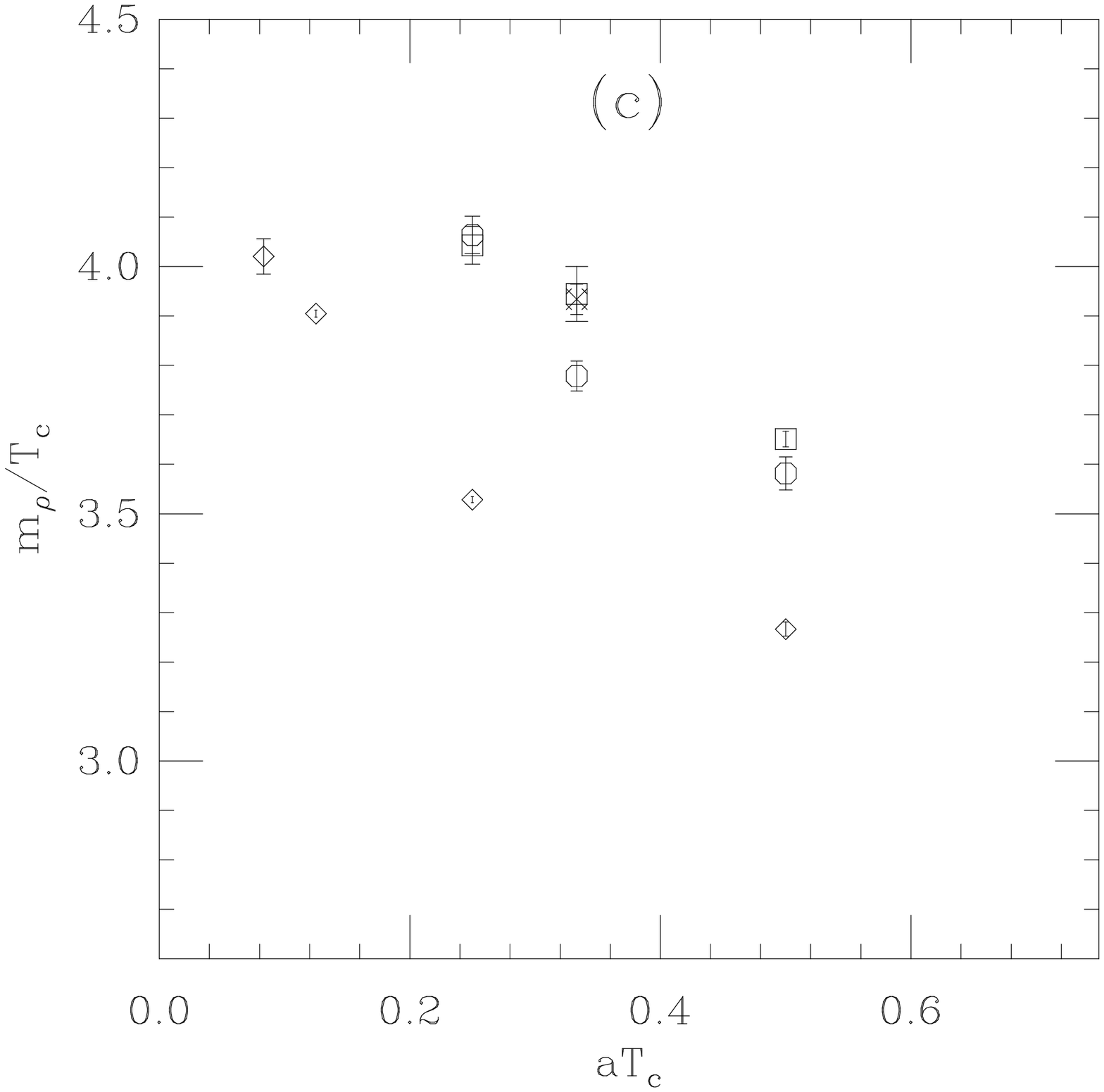}{80mm}
\ewxy{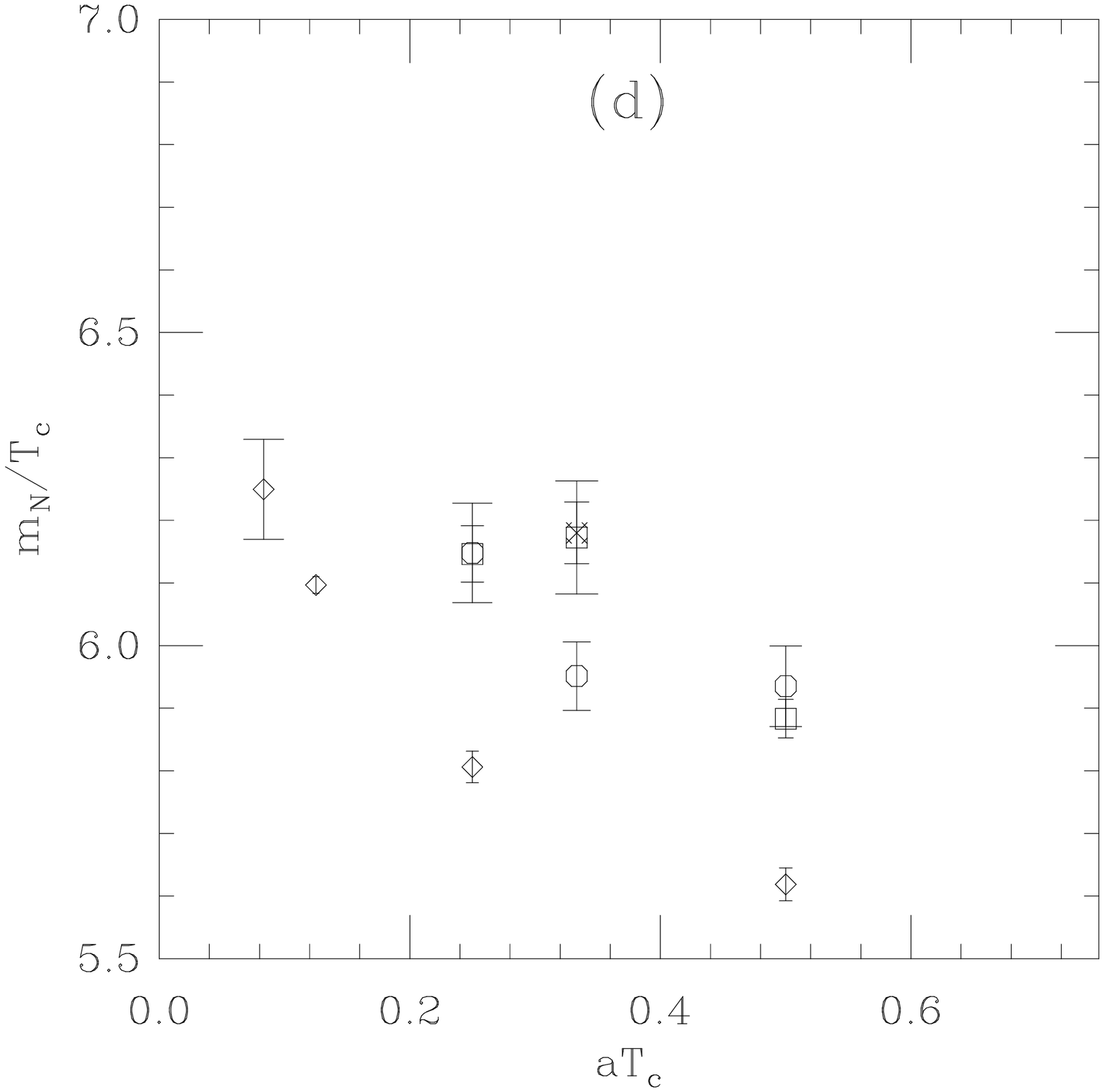}{80mm}}
\caption{Variation of  $m_\rho/T_c$ and $m_N/T_c$ vs. $aT_c$ 
at fixed $m_\pi/T_c= 2.75$ ((a) and (b)) and 3.15 ((c) and (d)),
for actions A (octagons), C (squares),  Wilson action (diamonds),
and TI clover action (fancy crosses).}
\label{fig:mvstc}
\end{figure}

Since the $\rho$ meson is the particle which shows the 
largest scaling violations, the best way to quantify improvement is
by taking ``sections'' of the $m_\rho/T_c$ and $m_N/T_c$ 
vs. $m_\pi/T_c$
plots and displaying them vs. $aT_c$ 
at fixed $m_\pi/T_c$
in Fig. \ref{fig:mvstc}.  The improved actions
at $aT_c=1/2$ seem to show the same level of scaling violations as Wilson
data at a lattice spacing a factor of 3 smaller. The smaller lattice
spacing data seem to pick up about a factor of two improvement in
lattice spacing, although the uncertainty in the data is larger.
Quenched calculations are thought to scale in difficulty like
$1/a^6$; the cost of action C is about a factor of 20 compared to
 the Wilson action.

My data by themselves do not suggest a unique way to extrapolate to
$a=0$.  FP actions are classically perfect with no $a^n$ scaling violations
for any $n$. Approximate FP actions generally have discretization
errors at all orders in $a$, though the coefficients of any order in
$a$ are typically much smaller than the corresponding coefficient
in an $a^N$ improved action (for $n>N)$. 

\section{Conclusions}

The important ingredients of these actions which contribute to their
improved scaling behavior are the hypercubic kinetic term and the
lattice anomalous magnetic moment term.  The  very small additive
renormalization of the bare lattice mass is due to the use of fermion-
gauge field couplings which are insensitive to the short-distance
fluctuations of the gauge fields.

The specific implementation of these ideas in the actions I have 
tested involves many arbitrary choices.  I believe
that essentially every choice I made for a particular parameterization could
be replaced by another choice, which would give an action which
would have the same quality of scaling
violations.  Some changes should add some additional good feature.
For example, it might be possible to find a parameterization of a fat link
which would lend itself to simple perturbation theory calculations.

There are several  obvious extensions of this work.  
The first involves the kinetic term:  It would be useful to
find a parameterization of the kinetic term which extends from zero mass to
very large mass. 
It would also be interesting to develop a fat link parameterization 
of the action which could be efficiently
incorporated into one of the standard algorithms
for dynamical fermions.
Next, is there a better parameterization of the anomalous magnetic moment 
term which might improve scaling?

Fixed point actions have many desirable formal properties \cite{PETER97}:
they include scale invariant instanton solutions,  the index theorem,
an absence of exceptional configurations, and a remnant of
chiral symmetry  \cite{GW,PLN}.
These properties may not be present in an action which is a bad approximation
to a FP action.  I do not know how well they are satisfied by these
actions (other than the apparent small renormalization of the quark mass.)

Indeed, the particular choice of a free field action which I made
was motivated only
by the locality and spectral properties of the free  action.
No attempt was made to optimize the chiral properties of the approximate action.
This is clearly the outstanding problem for future study.

\appendix
\setcounter{section}{0}
\setcounter{equation}{0}
\def\theequation{\Alph{section}.\arabic{equation}}

\section{Pure Gauge Actions}
This work used a new few-parameter approximation to an FP
action for $SU(3)$ gauge theory. Using it, isolated instanton configurations have constant actions to within 1.5 per cent.
 Like the action of Ref. \cite{SEPT},
it is a superposition of powers of the plaquette and the 
perimeter-6 ``twisted'' link $(x,y,z,-x,-y,-z)$. Like the
action of Ref. \cite{INST} it includes a constant term. It is designed
to be used for couplings such that the lattice spacing is $aT_c \simeq 1/3$ or
 1/4 to 1/8 or so. Explicitly
\bee
S(V) = c_0+{1 \over N_c} \sum_{\it C} ( c_1({\it C})(N_c-{\rm Tr}(V_{\it C})) +
                       c_2({\it C})(N_c-{\rm Tr}(V_{\it C}))^2+ ...
\label{ACTIONPAR}
\ee
with coefficients tabulated in Table \ref{tab:eightpar}.

\begin{table*}[hbt]
\caption{Couplings of the few-parameter approximate FP action.}
\label{tab:eightpar}
\begin{tabular*}{\textwidth}{@{}l@{\extracolsep{\fill}}lcccccc}
\hline
operator &  $c_1$ & $c_2$ & $c_3$ & $c_4$ & $c_0=-2.517$\\
 \hline
$c_{plaq}$ &  3.248  &  -1.580      & .1257 & .0576   \\
$c_{6-link}$ &  -.2810  &  .0051 &  .00490 & -.0096        \\
\hline
\end{tabular*}
\end{table*}

This action costs about a factor of 7 times the usual Wilson plaquette action
to simulate.

I have measured the critical couplings for the deconfinement transition
at $aT_c=1/2$, 1/3, and 1/4.  The critical couplings on the measured spatial
volumes and my extrapolation to infinite volume are shown 
in Table \ref{tab:betacrit}.  I have also measured the string tension
from Wilson loops at these values of the coupling, on $8^4$ lattices
(at $\beta=3.092$) and $12^4$ lattices for the other couplings.
The data was fitted to a static potential $V(r)$ of the form
\bee
   \label{eq:potfit}
V(r) = V_0 + \sigma r - E/r
\ee
 using the techniques of Ref. \cite{HELLV}.
The fit to the largest lattice spacing data is very difficult. The signal
from large $r$ is not good, and there is very little left of the
Coulomb part of the potential due to the coarseness of the lattice.
Nevertheless, I present the string tension and the Sommer
\cite{SOMMER} parameter $r_0$  ($r_0^2 dV(r_0)/dr= -1.65$) in Table
\ref{tab:betacrit}.  We see scaling within errors
for both these parameters (vs. $T_c$)
at $aT_c$ = 1/3 and 1/4. 
 There is a ten per cent scaling violation at $aT_c=1/2$.
The asymptotic value inferred from
large scale Wilson simulations \cite{BIELEFELD} is 
$\sqrt{\sigma}/T_c=1.600(11)$.

\begin{figure}[htb]
\begin{center}
\vskip -10mm
\leavevmode
\epsfxsize=60mm
\epsfbox[40 50 530 590]{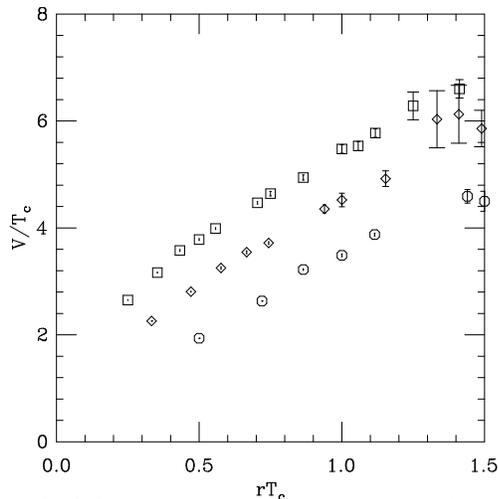}
\vskip -5mm
\end{center}
\caption{The potential of the approximate FP action, vs distance,
scaled with $T_c$.  Octagons show data for $aT_c=1/2$, 
diamonds for $aT_c=1/3$, and squares for $aT_c=1/4$.}
\label{fig:vtc234}
\end{figure}

Finally, I show a plot of $V(r)/T_c$ vs $rT_c$ for the three lattice
spacings in Fig. \ref{fig:vtc234}.  The overall vertical shift in the
 potentials is
not physical, but it allows the reader to separate the different data sets
by eye.

\begin{table*}[hbt]
\setlength{\tabcolsep}{1.5pc}
\caption{Critical couplings at finite volume and extrapolated to
infinite volume for the FP action with parameters in Table~1.}
\label{tab:betacrit}
\begin{tabular*}{\textwidth}{@{}l@{\extracolsep{\fill}}lcccccc}
\hline
volume   & $N_t=2$   & $N_t=3$ & $N_t=4$  \\
\hline
$4^3$    & 3.025(25)     &    &      \\
$6^3$    &  3.06(1)  & 3.47(1)   &     \\
$8^3$    &  3.08(1)    &  3.49(1)  &  3.67(1)   \\
$10^3$   & 3.085(5)    &  3.50(1)  &  3.69(1)   \\
$12^3$   &             &          &  3.69(1)    \\
infinite &  3.092(7)   &  3.50(1) &  3.70(1)   \\
\hline
$T_c/\Lambda$   & 8.67   & 8.96   & 8.33     \\
\hline
$a^2 \sigma$  & 0.56(5)  & .302(16)  & .164(3) \\
$\sqrt(\sigma)/T_c$ & 1.50(7)  &  1.65(2)  &  1.62(2)  \\
$r_0/a$       & 1.3(2)   & 2.19(2) & 2.93(1) \\
$r_0 T_c$     & .65(10)  & .730(7)  & .733(3)  \\
\hline
\end{tabular*}
\end{table*}

\section*{Acknowledgements}
This work is an outgrowth of an ongoing project 
with
 A.~Hasenfratz, P.~Hasenfratz, and F.~Niedermayer
to construct
 true FP actions for four-dimensional fermions. I am indebted to
them for many discussions about this work.
I would also like to thank
R.~Brower,
and
U.~Wiese
for valuable
conversations about FP actions, and S.~Gottlieb, U.~Heller,
R.~Sugar and D.~Toussaint for advice on supercomputing.
I  would like to thank   the Colorado high energy experimental
groups for allowing me to use their work stations.
This work was supported by the U.S. Department of 
Energy, with some computations done on the T3E at Pittsburgh Supercomputing
Center through resources awarded to the MILC collaboration.

\newcommand{\PL}[3]{{Phys. Lett.} {\bf #1} {(19#2)} #3}
\newcommand{\PR}[3]{{Phys. Rev.} {\bf #1} {(19#2)}  #3}
\newcommand{\NP}[3]{{Nucl. Phys.} {\bf #1} {(19#2)} #3}
\newcommand{\PRL}[3]{{Phys. Rev. Lett.} {\bf #1} {(19#2)} #3}
\newcommand{\PREPC}[3]{{Phys. Rep.} {\bf #1} {(19#2)}  #3}
\newcommand{\ZPHYS}[3]{{Z. Phys.} {\bf #1} {(19#2)} #3}
\newcommand{\ANN}[3]{{Ann. Phys. (N.Y.)} {\bf #1} {(19#2)} #3}
\newcommand{\HELV}[3]{{Helv. Phys. Acta} {\bf #1} {(19#2)} #3}
\newcommand{\NC}[3]{{Nuovo Cim.} {\bf #1} {(19#2)} #3}
\newcommand{\CMP}[3]{{Comm. Math. Phys.} {\bf #1} {(19#2)} #3}
\newcommand{\REVMP}[3]{{Rev. Mod. Phys.} {\bf #1} {(19#2)} #3}
\newcommand{\ADD}[3]{{\hspace{.1truecm}}{\bf #1} {(19#2)} #3}
\newcommand{\PA}[3] {{Physica} {\bf #1} {(19#2)} #3}
\newcommand{\JE}[3] {{JETP} {\bf #1} {(19#2)} #3}
\newcommand{\FS}[3] {{Nucl. Phys.} {\bf #1}{[FS#2]} {(19#2)} #3}

\end{document}